\documentclass[aps,prb,twocolumn,10pt,preprintnumbers,amsmath,amssymb,showkeys,floatfix]{revtex4-2}

\usepackage[T1]{fontenc}
\usepackage[utf8]{inputenc}
\usepackage{lmodern}
\usepackage{graphicx}
\graphicspath{{figures/}}
\usepackage{amsmath,amssymb,amsfonts}
\usepackage{physics}
\usepackage{hyperref}
\usepackage{xcolor}
\usepackage{bm}
\usepackage{mathtools}
\usepackage{booktabs}
\usepackage{orcidlink}
\usepackage{microtype}

\newcommand{\ii}{\mathrm{i}}
\newcommand{\bPi}{\bm{\Pi}}
\newcommand{\bp}{\bm{p}}
\newcommand{\bA}{\bm{A}}

\newcommand{\PhiAB}{\Phi_{\mathrm{AB}}}
\newcommand{\PhiQ}{\Phi_0}

\hypersetup{
  colorlinks=true,
  linkcolor=blue,
  citecolor=blue,
  urlcolor=blue,
  pdftitle={Ballistic transport and crystallographic control in altermagnetic quantum rings},
  pdfauthor={Edilberto O. Silva}
}

\begin{document}

\title{Ballistic transport and crystallographic control in altermagnetic quantum rings}

\author{Edilberto O. Silva\orcidlink{0000-0002-0297-5747}}
\email[Corresponding author: ]{edilberto.silva@ufma.br}
\affiliation{Programa de P\'os-Gradua\c{c}\~ao em F\'{\i}sica \& Coordena\c{c}\~ao do Curso de F\'isica -- Bacharelado, Universidade Federal do Maranh\~ao, 65080-805 S\~ao Lu\'{\i}s, Maranh\~ao, Brazil}

\date{October 19, 2026}

\begin{abstract}
We study coherent two-terminal transport through a finite-width $d$-wave altermagnetic quantum ring with nonmagnetic leads.  A confined Aharonov--Bohm flux remains exactly $h/e$ periodic and, in the symmetric one-mode geometry, produces an exact half-flux transmission zero.  A perpendicular field shifts the interference pattern according to an effective area $A_{\rm eff}=1.0342\,\pi r_0^2$ and the altermagnetic splitting resolves the common zero into nearby spin-dependent nodes.  Rotating the crystal reverses the transmitted spin, whereas the mirror orientation $\theta_A=\pi/4$ is exactly spin degenerate.  For a $50$~nm reference ring, sub-tesla fields yield windows with $|P_T|>0.99$, the most accessible of which requires a uniform field alone and no confined flux.  Finite-bias, thermal, disorder, Zeeman, and spin-conserving voltage-probe calculations show that the polarization is appreciably more stable than individual coherent-resonance amplitudes.  These results identify interference-node positions, flux periodicity, and crystal-controlled spin polarization as experimentally useful signatures of an altermagnetic mesoscopic interferometer.
\end{abstract}


\maketitle

\section{Introduction}
\label{sec:intro}

Altermagnets are compensated magnets whose crystal symmetries permit momentum-dependent spin splitting even without a net magnetization \cite{SmejkalPRX2022Phase,SmejkalPRX2022Landscape,JungwirthNatPhys2026,SongNatRevMat2025,JungwirthNature2026,McClartyRau2024,CheongClassification2025}.  Their spin texture may carry a $d$-wave form, with opposite spin sectors related by crystal rotation rather than simple momentum inversion.  Spectroscopic and transport evidence now spans several candidate materials and experimental probes \cite{KrempaskyNature2024,HarikiPRL2024,LeeMnTe2024,AminMnTe2024,FengRuO22022,ZhuPlaid2024,GuoRuO22024,GonzalezMnTe2024}, while recent theory has developed complementary descriptions of transport, nonlinear response, textures, and spin transfer \cite{SunPRB2023,ZarzuelaPRB2025,ZhangPRL2025,SooriPRM2026,GomonayTextures2024,SatoAHE2024,FangQuantumGeometry2024,JaeschkeSupercell2024,ZhangPairing2024}.  This makes phase-coherent mesoscopic structures a natural place to ask how the crystal anisotropy enters an actual scattering experiment.

Quantum rings are useful for this question because the magnetic flux enters directly as an interference phase.  Their AB response, phase rigidity, multichannel structure, resonant line shapes, and persistent-current physics are well established experimentally and theoretically \cite{AharonovBohm1959,TanInkson1996,ButtikerImryLandauer1985,TanInkson1999,YeyatiButtiker1995,HackenbroichWeidenmuller1996,Cernicchiaro1997,Aharony2006,Fuhrer2006,Nakanishi2004,Moldoveanu2005,Vargiamidis2006,FrustagliaRichter2004,Cheung1989,Levy1990,AmbegaokarEckern1990,Montambaux1990,vonOppen1991,Loss1990}.  We use the finite-width Tan--Inkson confinement because it keeps the radial problem controlled while allowing the magnetic field to penetrate the conducting annulus.  In an isolated circular ring a global rotation can remove the apparent dependence on the crystal angle.  Once leads are attached, that rotational freedom is lost: the contacts define the transport axis against which the altermagnetic crystal is measured.

For an experimental scale we take the layered vanadium oxytelluride $\mathrm{Rb}_{1-\delta}\mathrm{V}_2\mathrm{Te}_2\mathrm{O}$ as a reference. Spin-resolved photoemission resolves crystal-symmetry-paired opposite spin splittings in this metallic compound, and theory finds a quasi-two-dimensional spin-split Fermi surface with symmetry-controlled spin--momentum locking \cite{ZhangNatPhys2025RbV,HuPRX2025RbV,SongNatRevMat2025,JungwirthNature2026}. Its layered structure has also motivated first-principles tunnel-junction proposals \cite{MaoPRB2026RbV}. Neutron diffraction, however, reports a G-type antiferromagnetic structure below about $337$~K \cite{XieCPB2026RbV}. We use the compound only to set a realistic symmetry and device scale; the two-band Hamiltonian below is not presented as a first-principles fit to a specific Fermi-surface sheet.

This observation suggests a mesoscopic transport problem distinct from both conventional AB interferometry and bulk altermagnetic magnetotransport. The principal question is whether a compensated altermagnetic order can reshape coherent transmission through a ring in a controllable, orientation-dependent fashion. Closely related questions concern the fate of AB oscillations, the appearance of spin-resolved transmission imbalance, and the field-controlled phase and spin structure generated by a uniform orbital field. The detailed correspondence between individual closed-ring anticrossings and open-system resonances is left as a spectroscopy problem for future work.

We study this problem in a two-terminal geometry, starting from the continuum Hamiltonian and carrying the same model through a nonequilibrium Green's-function (NEGF) calculation \cite{Caroli1971,Datta1995}. The open contacts make the crystal angle observable and lead to simple angular identities: each spin sector is $\pi$ periodic, whereas the spin-summed conductance is $\pi/2$ periodic because a quarter-turn exchanges the two spin blocks. We treat a confined AB flux and a penetrating perpendicular field separately before combining them. This separation makes it possible to identify the nearly rigid field-induced translation of the AB pattern and the smaller corrections associated with the finite radial width of the ring. We then examine finite voltage and temperature windows, scalar disorder, and phase relaxation to see which spin-selective regimes survive outside the ideal coherent linear-response limit. Dimensional results are quoted in laboratory units for the reference device introduced below.

\section{Open altermagnetic quantum ring}
\label{sec:model}

\subsection{Central-region Hamiltonian}

We consider a two-dimensional annular conductor connected to a left ($L$) and a right ($R$) nonmagnetic lead. The central region is described by
\begin{equation}
H_C=H_0+H_{\mathrm{AM}},
\label{eq:HC}
\end{equation}
with
\begin{equation}
H_0=\frac{\bPi^2}{2m^*}+V_{\mathrm{TI}}(r),
\label{eq:H0}
\end{equation}
where $m^*$ is the effective mass and
\begin{equation}
\bPi=\bp+e\bA
\label{eq:kinetic_momentum}
\end{equation}
for an electron of charge $-e$. The finite-width Tan--Inkson confinement \cite{TanInkson1996} is
\begin{equation}
V_{\mathrm{TI}}(r)=\frac{a_1}{r^2}+a_2r^2-V_0,
\qquad
V_0=2\sqrt{a_1a_2},
\label{eq:TI}
\end{equation}
whose minimum occurs at
\begin{equation}
r_0=\left(\frac{a_1}{a_2}\right)^{1/4}.
\label{eq:r0}
\end{equation}
The additive constant in Eq.~\eqref{eq:TI} sets the minimum of the radial potential to zero.

The vector potential is written as the sum of a contribution from a perpendicular uniform field $B$ and a flux tube threading the inaccessible center of the ring,
\begin{equation}
\bA=\bA_B+\bA_{\mathrm{AB}}.
\label{eq:A_total}
\end{equation}
In the symmetric gauge,
\begin{equation}
\bA_B=\frac{Br}{2}\,\hat{\bm\phi},
\qquad
\bA_{\mathrm{AB}}=\frac{\PhiAB}{2\pi r}\,\hat{\bm\phi}.
\label{eq:A_symmetric}
\end{equation}
The distinction between $B$ and $\PhiAB$ is essential. The latter can be chosen to generate magnetic flux only through the excluded central region, whereas $B$ acts directly on the conducting annulus and therefore changes both interference phases and orbital dynamics.

For a minimal collinear $d$-wave altermagnet with conserved $\sigma_z$, we take
\begin{align}
H_{\mathrm{AM}}
={}&\frac{\beta}{2m^*}\Big[
\cos(2\theta_A)(\Pi_x^2-\Pi_y^2)
\nonumber\\
&\hspace{2.7cm}
+\sin(2\theta_A)\{\Pi_x,\Pi_y\}
\Big]\sigma_z,
\label{eq:HAM}
\end{align}
where $\beta$ is a dimensionless altermagnetic strength, $\theta_A$ measures the orientation of the crystal axes relative to the transport axis, and $\{A,B\}=AB+BA$. In the absence of additional spin-orbit or Zeeman terms, $[H_C,\sigma_z]=0$ and the problem decomposes into independent spin sectors $s=\pm1$.

For fixed $s$, Eqs.~\eqref{eq:H0} and \eqref{eq:HAM} give
\begin{align}
H_s={}&\frac{1}{2m^*}\Big[
(1+s\beta c_A)\Pi_x^2
+(1-s\beta c_A)\Pi_y^2
\nonumber\\
&\hspace{1.9cm}
+s\beta s_A\{\Pi_x,\Pi_y\}
\Big]+V_{\mathrm{TI}}(r),
\label{eq:Hs}
\end{align}
where $c_A=\cos(2\theta_A)$ and $s_A=\sin(2\theta_A)$. Equivalently, the kinetic energy is governed by the inverse-mass tensor
\begin{equation}
\mathsf{M}^{-1}_s
=\frac{1}{m^*}
\begin{pmatrix}
1+s\beta c_A & s\beta s_A\\
s\beta s_A & 1-s\beta c_A
\end{pmatrix}.
\label{eq:mass_tensor}
\end{equation}
Its principal values are $(1\pm\beta)/m^*$, independently of $s$. The continuum Hamiltonian is therefore elliptic and bounded from below for
\begin{equation}
|\beta|<1,
\label{eq:beta_bound}
\end{equation}
which will be taken as the natural parameter domain of the minimal effective-mass model.

\subsection{Leads and contact geometry}
\label{subsec:leads}

The left and right reservoirs are modeled as ideal spin-degenerate waveguides,
\begin{equation}
H_\alpha=\frac{\bPi^2}{2m_\alpha}+V_\alpha(y),
\qquad \alpha=L,R,
\label{eq:lead_hamiltonian}
\end{equation}
with spin-independent couplings $\tau_\alpha$ to the annular scattering region. The lead centers define the laboratory $x$ axis. In the simplest geometry they are attached at opposite points of the annulus near $(x,y)=(-r_0,0)$ and $(r_0,0)$, with equal widths and identical transverse confinement.

The production lattice uses an explicit open-contact construction.  Inside the horizontal contact aperture the Tan--Inkson confinement is set to zero once the neck leaves the nominal ring radius, while the altermagnetic hopping remains active through the finite coupling neck and terminates at the boundary to the ideal spin-degenerate lead self-energy.  Thus the reservoirs are nonmagnetic, but the central coupling region is part of the altermagnetic scatterer.  This is the implementation used for all figures below.

Because coherent resonance heights can depend on interface matching, we also tested a continuous taper of the Tan--Inkson confinement over $10$~nm and, separately, a smooth reduction of the altermagnetic anisotropy through the neck.  These interface variants change the absolute conductance substantially but preserve the large spin selectivity at the representative operating points (see Appendix~\ref{app:contact}).  We therefore treat contact geometry as a physical device parameter rather than as a numerical detail.

\subsection{Why the crystallographic angle becomes observable}
\label{subsec:theta_relevance}

For an isolated circular annulus, the radial confinement is invariant under arbitrary in-plane rotations. Let $\mathcal{R}(\varphi)$ denote the orbital rotation operator. The rotated altermagnetic form obeys
\begin{equation}
H_s(\theta_A)=\mathcal{R}(-\theta_A)
H_s(0)\mathcal{R}^{-1}(-\theta_A)
\label{eq:closed_rotation}
\end{equation}
when the entire system, including the gauge choice, is transformed consistently. The closed-ring spectrum is consequently independent of $\theta_A$.

In the open device, Eq.~\eqref{eq:closed_rotation} no longer represents a symmetry operation because the same rotation also moves the leads. Keeping the contacts fixed while rotating only the altermagnetic axes changes the relative orientation between the anisotropic spin-resolved dispersion and the injection direction. Thus $\theta_A$ becomes a genuine transport parameter. The angle is therefore redundant in the isolated circular problem but measurable once the contacts select a laboratory axis.

\section{Exact symmetry constraints}
\label{sec:symmetry}

Several exact identities follow directly from the minimal Hamiltonian and are useful for organizing the transport response.

\subsection{Angular periodicity and spin exchange}

Because the altermagnetic form depends on $2\theta_A$,
\begin{equation}
H_s(\theta_A+\pi)=H_s(\theta_A).
\label{eq:theta_pi}
\end{equation}
A quarter-turn gives a second identity,
\begin{equation}
H_s\left(\theta_A+\frac{\pi}{2}\right)
=H_{-s}(\theta_A),
\label{eq:spin_exchange}
\end{equation}
for fixed orbital fields, confinement, and contacts. Therefore, for spin-independent leads,
\begin{equation}
T_s\left(E;\theta_A+\frac{\pi}{2}\right)
=T_{-s}(E;\theta_A).
\label{eq:T_spin_exchange}
\end{equation}
The spin-summed transmission and conductance satisfy the exact relation
\begin{equation}
T\left(E;\theta_A+\frac{\pi}{2}\right)=T(E;\theta_A),
\qquad
G\left(\theta_A+\frac{\pi}{2}\right)=G(\theta_A),
\label{eq:G_pi2}
\end{equation}
whereas the transmission polarization
\begin{equation}
P_T=\frac{T_\uparrow-T_\downarrow}{T_\uparrow+T_\downarrow}
\label{eq:PT}
\end{equation}
changes sign,
\begin{equation}
P_T\left(\theta_A+\frac{\pi}{2}\right)=-P_T(\theta_A).
\label{eq:PT_pi2}
\end{equation}
Consequently, the leading angular harmonic of the spin-summed conductance is naturally compatible with $\cos(4\theta_A)$, even though the Hamiltonian itself contains $\cos(2\theta_A)$ and $\sin(2\theta_A)$. This distinction between spin-resolved and spin-summed angular periodicities is an important diagnostic of the model.

Changing the sign of $\beta$ also exchanges the two spin blocks,
\begin{equation}
H_s(-\beta)=H_{-s}(\beta),
\label{eq:beta_exchange}
\end{equation}
so the total conductance is even in $\beta$ while $P_T$ is odd in $\beta$.

\subsection{Mirror symmetry and the \texorpdfstring{$\theta_A=\pi/4$}{thetaA=pi/4} fixed point}

For the left--right device with mirror symmetry under $y\rightarrow-y$, reflection changes the sign of the mixed operator $\{\Pi_x,\Pi_y\}$ and reverses the orbital magnetic phases.  The spin-resolved transmission therefore obeys
\begin{equation}
T_s(\theta_A;B,\PhiAB)=T_s(-\theta_A;-B,-\PhiAB).
\label{eq:mirror_relation}
\end{equation}
Combining Eq.~\eqref{eq:mirror_relation} with the blockwise two-terminal reciprocity derived below gives $T_s(\theta_A)=T_s(-\theta_A)$ for the mirror-symmetric device.  At $\theta_A=\pi/4$, this evenness and the spin-exchange identity Eq.~\eqref{eq:T_spin_exchange} imply
\begin{equation}
T_\uparrow(\pi/4)=T_\downarrow(\pi/4),
\label{eq:pi4_degeneracy}
\end{equation}
for arbitrary $B$ and $\PhiAB$ within the minimal spin-conserving model.  The vanishing polarization at the mirror orientation is therefore an exact symmetry result, not a numerical coincidence.

\subsection{Flux periodicity and reciprocity}

For a pure AB flux localized in the excluded center, gauge invariance requires
\begin{equation}
G(\PhiAB+\PhiQ)=G(\PhiAB),
\qquad
\PhiQ=\frac{h}{e}.
\label{eq:AB_periodicity}
\end{equation}
This periodicity must not be imposed on the uniform-field problem, because a finite $B$ modifies the orbital motion throughout the conducting region.

The two-terminal conductance also obeys Onsager--Casimir reciprocity \cite{Onsager1931,Casimir1945,Buttiker1986}. For a magnetic order parameter that changes sign under time reversal,
\begin{equation}
G(B,\PhiAB,\beta)=G(-B,-\PhiAB,-\beta).
\label{eq:onsager_general}
\end{equation}
In the present spin-conserving model, the spin-summed conductance is even in $\beta$ by Eq.~\eqref{eq:beta_exchange}. Hence
\begin{equation}
G(B,\PhiAB,\beta)=G(-B,-\PhiAB,\beta)
\label{eq:onsager_total}
\end{equation}
for ideal nonmagnetic contacts and in the absence of additional spin-selective terms. In particular, $G(\PhiAB)=G(-\PhiAB)$ in the pure-AB two-terminal geometry and $G(B)=G(-B)$ when $\PhiAB=0$. Violations of these identities would signal either a numerical inconsistency or additional physics outside the minimal Hamiltonian.

The statement is stronger in the present spin-conserving model. At fixed $s$, the scalar lattice Hamiltonian has real coefficients before the electromagnetic links are inserted, so complex conjugation reverses all orbital Peierls phases,
\begin{equation}
H_s(B,\PhiAB)^*=H_s(-B,-\PhiAB).
\label{eq:block_flux_conjugation}
\end{equation}
Two-terminal reciprocity and unitarity therefore imply the blockwise identity
\begin{equation}
T_s(B,\PhiAB)=T_s(-B,-\PhiAB),
\label{eq:spin_phase_rigidity}
\end{equation}
for each spin sector separately. Thus $T_s(B)=T_s(-B)$ in the pure uniform-field problem, while the pure-AB response is even in $\PhiAB$. Together with flux-quantum periodicity in the latter case, the linear-response conductance can be expanded as
\begin{equation}
G(\PhiAB)=G_0+\sum_{n\ge1}c_n\cos\!\left(2\pi n\frac{\PhiAB}{\PhiQ}\right),
\label{eq:phase_rigid_fourier}
\end{equation}
without sine harmonics.  This is the usual two-terminal AB phase rigidity \cite{YacobyPRB1996,Buttiker1986}: changing $\theta_A$ may redistribute the cosine amplitudes, enhance higher harmonics, or produce discrete $\pi$ inversions when a coefficient changes sign, but it cannot generate an arbitrary continuous shift of the linear-response AB phase.

\section{Landauer--B\"uttiker and NEGF formulation}
\label{sec:transport}

\subsection{Gauge covariance and conservation laws}

The transport observables must be independent of the gauge used for the vector potential. For an electron of charge $-e$, a gauge change $\bA\rightarrow\bA+\bm{\nabla}\chi$ is accompanied by
\begin{equation}
U_\chi(\bm r)=\exp\!\left[-\ii\frac{e}{\hbar}\chi(\bm r)\right],
\qquad
H_s' = U_\chi H_s U_\chi^\dagger .
\label{eq:gauge_unitary}
\end{equation}
On the lattice this is the usual endpoint rephasing of every Peierls link.  If the same transformation is applied consistently to the central region and lead-contact subspaces, then
\begin{equation}
\begin{aligned}
G_s^{r\prime}&=U_\chi G_s^r U_\chi^\dagger,\\
\Sigma_{\alpha,s}^{r\prime}&=U_\chi\Sigma_{\alpha,s}^rU_\chi^\dagger,\\
\Gamma_{\alpha,s}'&=U_\chi\Gamma_{\alpha,s}U_\chi^\dagger .
\end{aligned}
\label{eq:gauge_green}
\end{equation}
The Caroli trace \cite{Caroli1971} is therefore unchanged by cyclicity,
\begin{equation}
\Tr\!\left[\Gamma_L'G^{r\prime}\Gamma_R'G^{a\prime}\right]
=\Tr\!\left[\Gamma_LG^r\Gamma_RG^a\right].
\label{eq:gauge_caroli}
\end{equation}
Moving the singular AB branch cut without changing the enclosed flux is the discrete version of the same gauge transformation.  These relations are the analytic reason why the calculated transmission cannot depend on the Landau-gauge origin or on the chosen branch cut \cite{BarangerStone1989,Datta1995,Hofstadter1976,DoPeierls2022}.

For the static coherent problem the scattering matrix is unitary.  For each conserved spin block and each energy,
\begin{equation}
R_s(E)+T_s(E)=N_s(E),
\label{eq:current_conservation_scattering}
\end{equation}
where $N_s$ is the number of incoming propagating modes.  In NEGF language the same statement follows from
\begin{equation}
G_s^r-G_s^a=-\ii G_s^r(\Gamma_{L,s}+\Gamma_{R,s})G_s^a,
\label{eq:green_spectral_identity}
\end{equation}
which is the open-system continuity identity behind the Landauer current formula \cite{FisherLee1981,Datta1995,Beenakker1997}.  Because the minimal Hamiltonian obeys $[H,\sigma_z]=0$, this conservation law applies separately to $s=\uparrow,\downarrow$.  Separate spin-current conservation is therefore a property of the present spin-conserving model; spin--orbit coupling or explicit spin-flip scattering would mix the two blocks.

The coherent Hamiltonian is time independent, so scattering is elastic and energy is conserved at a given incident energy.  The voltage-probe calculation used later is different: imposing $I_{p,s}=0$ guarantees particle-current conservation at every fictitious reservoir, but a voltage probe can exchange energy with the conductor.  We therefore use it only as a phenomenological model of phase relaxation, not as a microscopic energy-conserving scattering mechanism \cite{ButtikerPRB1986Coherence,DAmatoPastawski1990,Forster2007}.

\subsection{Linear-response conductance}

At zero temperature and infinitesimal bias, the two-terminal Landauer conductance is \cite{Buttiker1986,ButtikerImryLandauer1985,SivanImry1986,BarangerStone1989,Datta1995,Beenakker1997}
\begin{equation}
G(E_F)=\frac{e^2}{h}\sum_s T_s(E_F),
\label{eq:Landauer}
\end{equation}
where
\begin{equation}
T_s(E)=\sum_{n\in L}\sum_{m\in R}T^{(s)}_{mn}(E)
\label{eq:channel_sum}
\end{equation}
contains all propagating transverse channels. The transmission eigenvalues must satisfy
\begin{equation}
0\le \tau_{n,s}\le 1,
\label{eq:eigenvalue_bound}
\end{equation}
while current conservation requires equality between the total incoming probability and the sum of reflected and transmitted probabilities for each open channel.

At finite temperature,
\begin{equation}
G(T)=\frac{e^2}{h}\sum_s\int\dd E\,
T_s(E)\left(-\frac{\partial f}{\partial E}\right),
\label{eq:finiteT}
\end{equation}
which makes thermal averaging of narrow resonances explicit.

\subsection{Green functions and lead self-energies}

For numerical work we use the retarded Green function of the central region,
\begin{equation}
G_s^r(E)=
\left[(E+\ii\eta)I-H_{C,s}-\Sigma_{L,s}^r(E)-\Sigma_{R,s}^r(E)\right]^{-1}.
\label{eq:Gr}
\end{equation}
The lead self-energy is
\begin{equation}
\Sigma_{\alpha,s}^r=\tau_\alpha^\dagger g_{\alpha}^r\tau_\alpha,
\label{eq:selfenergy}
\end{equation}
where $g_\alpha^r$ is the surface Green function of the semi-infinite lead \cite{LopezSancho1985}. Since the leads and contacts are initially spin independent, the explicit spin index on $g_\alpha^r$ can be omitted. The level-broadening matrices are
\begin{equation}
\Gamma_{\alpha,s}=\ii\left(\Sigma_{\alpha,s}^r-\Sigma_{\alpha,s}^a\right),
\label{eq:Gamma}
\end{equation}
and the Caroli transmission formula \cite{Caroli1971,MeirWingreen1992} reads
\begin{equation}
T_s(E)=\Tr\left[
\Gamma_{L,s}G_s^r\Gamma_{R,s}G_s^a
\right].
\label{eq:Caroli}
\end{equation}
The equivalence between scattering-matrix and Green-function descriptions provides an independent numerical cross-check \cite{FisherLee1981,Datta1995}.

\subsection{Finite-bias current and two-terminal resistance}

For electrochemical potentials $\mu_L$ and $\mu_R$, the coherent current is
\begin{equation}
I_s=\frac{e}{h}\int\dd E\,
T_s(E)\left[f_L(E)-f_R(E)\right],
\label{eq:current}
\end{equation}
with $I=I_\uparrow+I_\downarrow$. A spin polarization of the transport current can be defined as
\begin{equation}
P_I=\frac{I_\uparrow-I_\downarrow}{I_\uparrow+I_\downarrow},
\label{eq:PI}
\end{equation}
provided the denominator is nonzero. For sufficiently small bias, Eq.~\eqref{eq:current} reduces to $I\simeq GV$.

The two-terminal resistance is $R=1/G$ whenever $G$ is finite.  We use it only as a complementary representation of interference minima; no bulk magnetoresistance coefficient is inferred from a two-terminal resonance whose depth is discretization- and contact-sensitive.

\section{Gauge-covariant lattice discretization}
\label{sec:lattice}

A reliable lattice representation is essential because an arbitrary $\theta_A$ generates the mixed operator $\{\Pi_x,\Pi_y\}$. We discretize the central region on a square finite-difference grid of physical spacing $\Delta$ and define the kinetic energy associated with that spacing,
\begin{equation}
t_\Delta=\frac{\hbar^2}{2m^*\Delta^2}.
\label{eq:tDelta}
\end{equation}
For a directed nearest-neighbor bond $i\rightarrow j$, the Peierls link \cite{Peierls1933,Hofstadter1976} is
\begin{equation}
U_{ij}=
\exp\left[
\ii\frac{e}{\hbar}\int_{\bm r_i}^{\bm r_j}\bA\cdot\dd\bm\ell
\right],
\qquad U_{ji}=U_{ij}^*.
\label{eq:link}
\end{equation}
This convention follows directly from $\bPi=-\ii\hbar(\bm\nabla+\ii e\bA/\hbar)$.

\subsection{Axial hoppings}

For spin $s$, the $x$ and $y$ nearest-neighbor amplitudes are
\begin{align}
t_{x,s}&=t_\Delta\left(1+s\beta\cos2\theta_A\right),
\label{eq:tx}\\
t_{y,s}&=t_\Delta\left(1-s\beta\cos2\theta_A\right).
\label{eq:ty}
\end{align}
The corresponding matrix elements are
\begin{align}
(H_s)_{i,i\pm\hat{x}}&=-t_{x,s}U_{i,i\pm\hat{x}},
\label{eq:hx}\\
(H_s)_{i,i\pm\hat{y}}&=-t_{y,s}U_{i,i\pm\hat{y}}.
\label{eq:hy}
\end{align}
Their diagonal kinetic contribution is
\begin{equation}
2t_{x,s}+2t_{y,s}=4t_\Delta,
\label{eq:onsite_kinetic}
\end{equation}
which is independent of $s$, $\beta$, and $\theta_A$. Thus
\begin{equation}
(H_s)_{ii}=4t_\Delta+V_{\mathrm{TI}}(r_i)+V_{\mathrm{match}}(\bm r_i),
\label{eq:onsite}
\end{equation}
where $V_{\mathrm{match}}$ denotes any smooth contact-matching potential.

\subsection{Mixed derivative and diagonal hoppings}

Following the Peierls construction for magnetic tight-binding models \cite{Hofstadter1976,DoPeierls2022}, and to discretize the mixed term without sacrificing gauge covariance, we use a symmetrized two-link transporter between diagonal sites. For example,
\begin{align}
W_i^{++}=\frac{1}{2}\Big[
U_{i,i+\hat{x}}U_{i+\hat{x},i+\hat{x}+\hat{y}}
+&U_{i,i+\hat{y}}U_{i+\hat{y},i+\hat{x}+\hat{y}}
\Big],
\label{eq:Wpp}
\end{align}
with analogous definitions $W_i^{+-}$, $W_i^{-+}$, and $W_i^{--}$. These objects average the two shortest paths connecting a site to a diagonal neighbor and transform covariantly under a lattice gauge transformation.

The centered approximation to the symmetrized covariant derivative is
\begin{align}
\mathcal{D}_{xy}\psi_i=\frac{1}{4\Delta^2}\Big(&
W_i^{++}\psi_{i+\hat{x}+\hat{y}}
-W_i^{+-}\psi_{i+\hat{x}-\hat{y}}
\nonumber\\
&-W_i^{-+}\psi_{i-\hat{x}+\hat{y}}
+W_i^{--}\psi_{i-\hat{x}-\hat{y}}
\Big),
\label{eq:Dxy}
\end{align}
which approaches $\tfrac12(D_xD_y+D_yD_x)$ in the continuum limit. Since
\begin{equation}
\{\Pi_x,\Pi_y\}=-2\hbar^2\mathcal{D}_{xy}+\mathcal{O}(\Delta^2),
\label{eq:mixed_continuum}
\end{equation}
the mixed altermagnetic term produces diagonal hoppings
\begin{align}
(H_s)_{i,i+\hat{x}+\hat{y}}
&=-\frac{s\beta\sin2\theta_A}{2}t_\Delta W_i^{++},
\label{eq:hpp}\\
(H_s)_{i,i+\hat{x}-\hat{y}}
&=+\frac{s\beta\sin2\theta_A}{2}t_\Delta W_i^{+-},
\label{eq:hpm}
\end{align}
with the same sign pattern for the opposite diagonals and Hermitian-conjugate matrix elements for reversed bonds. At $B=\PhiAB=0$, Eqs.~\eqref{eq:hpp}--\eqref{eq:hpm} reduce to the standard central-difference stencil for $\partial_x\partial_y$.

The path-symmetrized construction is preferable to assigning a single diagonal Peierls phase, because the latter chooses one of two inequivalent paths in a finite magnetic field and can introduce avoidable $\mathcal{O}(\Delta^2)$ gauge artifacts. Hermiticity should be checked numerically through $\|H-H^\dagger\|$ before any transport calculation.

\subsection{Treatment of uniform and AB fluxes}

For a uniform field and leads oriented along $x$, we use the shifted Landau gauge
\begin{equation}
\bA_B=-B(y-y_g)\,\hat{\bm x},
\label{eq:landau_shift}
\end{equation}
where changing $y_g$ is a pure gauge transformation. This choice preserves translation invariance along the ideal leads. The field is included in the leads as well as in the central region, so the lead surface Green functions and self-energies are recomputed at every $B$ rather than reusing their zero-field values. In a principal-layer representation, the transverse onsite block is field independent in this gauge, whereas the longitudinal hopping matrix acquires the $y$-dependent Peierls phase. We obtain the retarded surface Green function by iterative decimation of the semi-infinite block-tridiagonal lead \cite{LopezSancho1985}. In the $B\rightarrow0$ limit, this construction reproduces the analytic transverse-mode surface Green function used in the zero-field construction.

A pure AB flux can be implemented by a gauge in which the accumulated phase is concentrated on a branch cut inside the finite scattering region. In this representation the semi-infinite leads remain strictly periodic, while the total phase around any closed loop encircling the hole is
\begin{equation}
\sum_{\mathcal{C}}\arg U_{ij}
=2\pi\frac{\PhiAB}{\PhiQ}
\quad (\mathrm{mod}\;2\pi).
\label{eq:loop_phase}
\end{equation}
Numerical observables must be unchanged if the branch cut or gauge is moved while the physical flux is kept fixed.

\section{Experimental reference and physical parametrization}
\label{sec:SIreference}

The continuum Hamiltonian is generic. To report the transport results in directly measurable units we define a reference mesoscopic device motivated by the layered $d$-wave spin-split electronic structure of $\mathrm{Rb}_{1-\delta}\mathrm{V}_2\mathrm{Te}_2\mathrm{O}$ \cite{ZhangNatPhys2025RbV,HuPRX2025RbV,SongNatRevMat2025,JungwirthNature2026}. We take
\begin{equation}
 m^*=0.30m_e=2.733\times10^{-31}\ \mathrm{kg},
 \qquad
 r_0=50\ \mathrm{nm},
 \label{eq:SI_mass_radius}
\end{equation}
with numerical radial truncations at $r_{\rm in}=25$~nm and $r_{\rm out}=85$~nm and a nominal lead width of $30$~nm. The Fermi level is placed $4.064$~meV above the selected effective-band edge and the calculations use $\beta=0.30$ unless stated otherwise. The reference mass is a transparent effective-device parameter, not an experimentally extracted transport mass for this compound; all energy, voltage, temperature, and current scales can therefore be rescaled immediately if a different low-energy mass is established for a chosen Fermi-surface pocket.

The Tan--Inkson potential is specified in SI units by
\begin{align}
 a_1&=1.0174\times10^{-35}\ \mathrm{J\,m^2},\\
 a_2&=1.6278\times10^{-6}\ \mathrm{J\,m^{-2}},\\
 V_0&=8.1390\times10^{-21}\ \mathrm{J}=50.80\ \mathrm{meV},
\label{eq:SI_TI_parameters}
\end{align}
which places its minimum at $r_0=50$~nm. All production curves, including the pure-AB data, use the same finite-difference spacing $\Delta=2.14$~nm.  Finer meshes down to $1.00$~nm are used only for convergence checks. These mesh lengths are numerical coarse-graining scales of the effective-mass envelope theory, not atomic lattice constants.

For reference, the magnetic flux quantum is
\begin{equation}
 \Phi_0=\frac{h}{e}=4.13567\times10^{-15}\ \mathrm{Wb}=4.13567\ \mathrm{fWb},
\label{eq:SI_fluxquantum}
\end{equation}
and one flux quantum through the mean-radius circle corresponds to $B=0.52657$~T. The conductance quantum is $e^2/h=38.7405~\mu$S. The low-energy splitting generated by $\beta=0.30$ is of order a few meV in the transport window, deliberately much smaller than the large momentum-dependent splittings observed across the full experimental band structure \cite{ZhangNatPhys2025RbV}; the calculation should therefore be read as a conservative low-energy interferometer model rather than a global band fit.

\section{Numerical implementation}
\label{sec:numerics}

The central Hamiltonian is assembled as a sparse gauge-covariant finite-difference operator and coupled to semi-infinite nonmagnetic leads through retarded self-energies. Zero-field lead surface Green functions are evaluated in the transverse-mode basis \cite{Datta1995}, while magnetic leads are treated by iterative decimation \cite{LopezSancho1985}. Spin-resolved transmissions are obtained from the Caroli formula \cite{Caroli1971}, and Fisher--Lee amplitudes \cite{FisherLee1981} are used as an independent internal current-conservation check. The confined AB flux is implemented through a singular-gauge branch cut that does not intersect the leads; the uniform field is included in both the central region and the leads. Gauge shifts, branch-cut moves, lattice refinement, lead-width variation, Hermiticity, and scattering unitarity were verified internally, but these validation diagnostics are not used as physics figures in the main text.

The open device has one propagating mode per spin at $E_F=4.064$~meV for the final reference geometry. All figures and quoted results below are expressed in physical units: magnetic field in tesla, confined flux in webers (displayed as fWb), energy in meV, voltage in mV, current in nA, conductance in $\mu$S, and temperature in kelvin.

The narrow coherent resonances converge more slowly in amplitude than the symmetry-protected relations and spin polarization.  Refinement from $\Delta=2.14$~nm to $1.00$~nm leaves the sign and near-unity magnitude of $P_T$ stable at the principal operating points but can shift the absolute resonant conductance by tens of percent.  Consequently, conductances, currents, and resistances below are representative device values quoted to modest precision rather than continuum-extrapolated material constants.  The full convergence table and interface tests are provided in Appendices~\ref{app:convergence} and \ref{app:contact}; no convergence or error plots are used as physics figures.

\section{Confined Aharonov--Bohm transport}
\label{sec:ab}
\label{sec:pureAB}

We first set the uniform magnetic field to zero and retain only a confined flux through the inaccessible center of the annulus. The reference geometry has $r_0=50$~nm, $r_{\rm in}=25$~nm, $r_{\rm out}=85$~nm, a $30$~nm lead width, $E_F=4.064$~meV, and $\beta=0.30$. The pure-AB curves use the same $2.14$~nm finite-difference spacing as the remaining production calculations. The flux is implemented in a singular gauge: every hopping crossing a branch cut from the excluded central region to the exterior acquires the winding phase required by $\Phi_{\rm AB}$, whereas contractible loops enclose no lattice field.

\begin{figure*}[t]
\centering
\includegraphics[width=\textwidth]{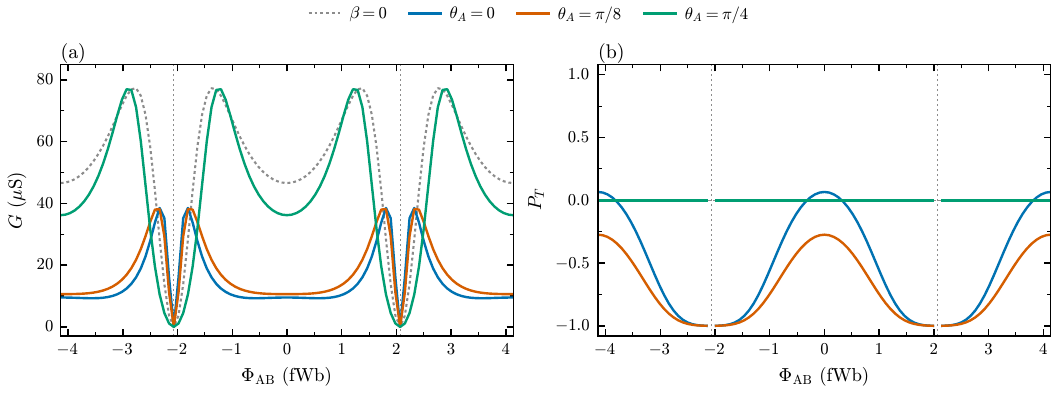}
\caption{Confined Aharonov--Bohm transport for a $50$~nm mean-radius ring at $E_F=4.064$~meV and $\beta=0.30$. (a) Total conductance for three crystallographic orientations; the dotted curve is the conventional $\beta=0$ ring. The response is even in confined flux and periodic with $\Phi_0=4.13567$~fWb. (b) Transport spin polarization. Points at the exact half-flux transmission zeros are omitted because the ratio defining $P_T$ is then indeterminate. The mirror orientation $\theta_A=\pi/4$ is spin degenerate.}
\label{fig:ab_interference}
\end{figure*}

Figure~\ref{fig:ab_interference} shows two AB periods, every plotted flux point being evaluated directly rather than reconstructed from a single half period. The response repeats after $\Phi_0=4.13567$~fWb and is even under $\Phi_{\rm AB}\rightarrow-\Phi_{\rm AB}$, as required for a two-terminal coherent conductor at $B=0$.  Both relations are measured rather than imposed: over the sampled range they hold to better than $3\times10^{-14}\,e^2/h$ at every $\theta_A$, which is the solver floor rather than a physical residual. The crystallographic angle changes the harmonic amplitudes without generating a continuous AB phase shift. A discrete Fourier analysis confirms that the sine components remain numerically negligible in this limit while the cosine harmonics vary strongly with $\theta_A$.

A symmetric one-mode device develops a deep destructive-interference minimum at
\begin{equation}
 \Phi_{\rm AB}=\frac{\Phi_0}{2}=2.06783\ \mathrm{fWb}.
\label{eq:half_flux_node}
\end{equation}
For the three orientations shown in Fig.~\ref{fig:ab_interference}, the residual conductance at this point lies below the numerical plotting scale.  In the one-mode symmetric device the node is exact.  Let $z=\exp(2\pi i\Phi_{\rm AB}/\Phi_0)$ and write the left-to-right transmission amplitude as a winding expansion $\mathcal A_{LR}(z)=\sum_w a_w z^w$.  A $C_2$ rotation maps a left-to-right path of winding $w$ to a right-to-left path with winding $w+1$, so $\mathcal A_{RL}(z)=z\mathcal A_{LR}(z)$.  Two-terminal reciprocity at $B=0$ \cite{Buttiker1986} gives $\mathcal A_{RL}(z)=\mathcal A_{LR}(z^{-1})$.  At half flux, $z=z^{-1}=-1$, and hence
\begin{equation}
\mathcal A_{LR}(-1)=-\mathcal A_{LR}(-1)=0.
\label{eq:half_flux_proof}
\end{equation}
The proof applies for arbitrary $\beta$ and $\theta_A$ as long as the $C_2$ geometry and one propagating channel are retained.  The relevant perturbations are those that break the $C_2$ geometry: a scalar onsite profile odd in $x$ lifts the zero quadratically in its amplitude, whereas a $C_2$-preserving profile leaves it intact even when orders of magnitude larger.  Scalar disorder, additional propagating modes, or a penetrating $B$ field likewise remove the common-spin cancellation, while the mirror symmetry $y\rightarrow-y$ is not required for it.

The spin polarization can be large away from the node. At $\theta_A=\pi/8$ and $\Phi_{\rm AB}=1.551$~fWb the reference device gives
\begin{equation}
T_\uparrow\simeq0.0110,\qquad T_\downarrow\simeq0.747,
\label{eq:ab_spin_example}
\end{equation}
corresponding to $G\simeq29~\mu$S and $P_T\simeq-0.971$ on the common $2.14$~nm grid. The precise spin preference is contact dependent, so this result identifies an accessible interferometric spin-filter regime rather than a universal bulk polarization.

For the mirror-symmetric geometry, $\theta_A=\pi/4$ yields $T_\uparrow=T_\downarrow$ and therefore $P_T=0$ over the complete pure-AB period. This follows directly from Eqs.~\eqref{eq:mirror_relation} and \eqref{eq:pi4_degeneracy}.  In Fig.~\ref{fig:ab_interference} the spin-summed $\theta_A=\pi/4$ trace also lies close to the conventional $\beta=0$ ring at $B=0$: here $\cos 2\theta_A=0$, so the axial hoppings recover their isotropic values and the altermagnetic correction is carried entirely by the mixed diagonal hopping.

\section{Uniform-field magnetotransport}
\label{sec:uniformB}

We next set $\Phi_{\rm AB}=0$ and allow a uniform perpendicular field to penetrate both the annulus and the ideal leads.  Trajectories across the finite radial width enclose a distribution of areas, so the relevant interferometric area need not coincide exactly with the nominal mean-radius value $\pi r_0^2$.  The final calculations use the $2.14$~nm grid and the same $E_F=4.064$~meV, $r_0=50$~nm, and $\beta=0.30$ reference device.

\subsection{Effective-area quasi-periodicity}

\begin{figure*}[t]
\centering
\includegraphics[width=\textwidth]{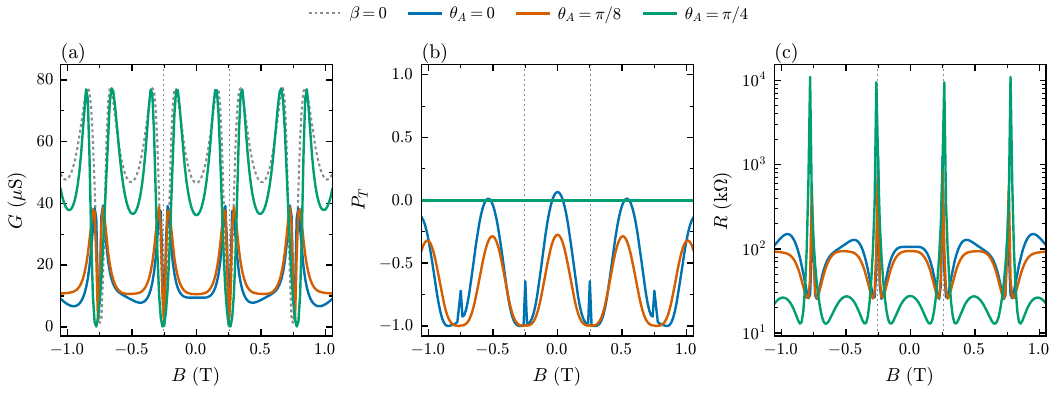}
\caption{Uniform-field magnetotransport of the $50$~nm reference ring at $E_F=4.064$~meV and $\beta=0.30$. (a) Total conductance for three crystallographic orientations; the dotted curve is the conventional $\beta=0$ ring. The response is even in field and displays a quasi-period set by the effective enclosed area. (b) Transport spin polarization. The mirror-symmetric orientation $\theta_A=\pi/4$ remains spin degenerate. (c) Two-terminal resistance on a logarithmic scale. Large peaks occur when the field drives the device close to a coherent transmission minimum.}
\label{fig:uniform_magnetotransport}
\end{figure*}

Figure~\ref{fig:uniform_magnetotransport} shows an even-in-$B$ magnetoconductance with narrow coherent resonances and an approximately periodic envelope.  The appropriate period is set by the effective enclosed area extracted independently from the combined-field AB phase,
\begin{equation}
A_{\rm eff}=\frac{\Phi_0}{2\pi}\frac{d\delta_1}{dB}
=8.12\times10^{-15}~\mathrm{m^2}
=1.0342\,\pi r_0^2,
\label{eq:Aeff}
\end{equation}
which gives
\begin{equation}
\Delta B_{\rm eff}=\frac{\Phi_0}{A_{\rm eff}}=0.509~\mathrm{T}.
\label{eq:Beff_period}
\end{equation}
Using $\Phi_0/(\pi r_0^2)=0.5266$~T instead would compare points displaced by about $17$~mT from the actual interference period, enough to move between a narrow valley and a peak.  The small residual mismatch after shifting by $\Delta B_{\rm eff}$ is the physically relevant finite-width orbital correction rather than a dominant loss of AB periodicity.

The few-percent excess over the nominal area follows from the radial extent of the Tan--Inkson ground state.  Expanding the confinement about its minimum gives $V''(r_0)=6a_1/r_0^4+2a_2=8a_2$ and therefore a radial level spacing
\begin{equation}
\hbar\omega_r=\hbar\sqrt{8a_2/m^*}=4.544~\mathrm{meV},
\label{eq:radial_spacing}
\end{equation}
which exceeds $E_F=4.064$~meV and so confirms analytically the single-propagating-mode regime assumed throughout.  The corresponding zero-point spread is $\langle(r-r_0)^2\rangle=\hbar/(2m^*\omega_r)=2.795\times10^{-17}$~m$^2$, an rms radial width of $5.29$~nm, so that
\begin{equation}
\frac{\langle r^2\rangle}{r_0^2}
=1+\frac{\langle(r-r_0)^2\rangle}{r_0^2}
=1.0112 .
\label{eq:area_harmonic}
\end{equation}
Including the leading anharmonicity of $V_{\rm TI}$, whose cubic coefficient $V'''(r_0)=-24a_1/r_0^5$ displaces the radial expectation value outward by $\langle r-r_0\rangle=+0.84$~nm, raises the estimate to $1.0447$.  The measured $A_{\rm eff}=1.0342\,\pi r_0^2$ falls between these two limits, so the effective interferometric area is accounted for by the finite radial extent of the transmitted state rather than being an empirical fit parameter.  Since $\langle(r-r_0)^2\rangle\propto a_2^{-1/2}$, a stiffer radial confinement should push $A_{\rm eff}$ back toward $\pi r_0^2$.

The nominal mean-area half-flux field,
\begin{equation}
 B_{1/2}^{\rm nom}=\frac{\Phi_0}{2\pi r_0^2}=0.26328~\mathrm{T},
\label{eq:uniform_half_field}
\end{equation}
is not the actual destructive-interference condition once the effective area is taken into account.  The rigid-shift estimate instead gives
\begin{equation}
 B_{1/2}^{\rm eff}=\frac{\Phi_0}{2A_{\rm eff}}=0.2546~\mathrm{T}.
\label{eq:uniform_half_effective}
\end{equation}
A bounded minimization of the total transmission on the production grid at $\theta_A=\pi/8$ places the conductance minimum at $B=0.253229$~T with $G_{\min}=8.60\times10^{-4}~\mu$S.  Thus the penetrating field primarily \emph{shifts} the half-flux cancellation rather than lifting it at the nominal mean-area field.

The altermagnetic term adds a second, more diagnostic effect: the common pure-AB zero resolves into nearby spin-dependent transmission nodes,
\begin{align}
 B_{\downarrow}&=0.253226~\mathrm{T},\\
 B_{\uparrow}&=0.256108~\mathrm{T},
\label{eq:spin_nodes}
\end{align}
on the $\Delta=2.14$~nm production grid.  Their separation is only a few millitesla and is more mesh sensitive than their individual positions, so we do not treat the numerical value of $\Delta B=B_\uparrow-B_\downarrow$ as a continuum prediction.  The mean position,
\begin{equation}
 \bar B=\frac{B_\uparrow+B_\downarrow}{2}=0.254667~\mathrm{T},
\end{equation}
provides an independent interferometric-area measurement,
\begin{equation}
 \frac{A_{\rm eff}^{(\rm node)}}{\pi r_0^2}
 =\frac{\Phi_0}{2\bar B\,\pi r_0^2}=1.03384,
\label{eq:Aeff_nodes}
\end{equation}
which agrees within $0.04\%$ with the Fourier-phase value $1.03421$ in Eq.~\eqref{eq:Aeff}.  This agreement between a phase observable and the positions of two transmission zeros is a stringent internal consistency check.  Upon refinement from $\Delta=2.14$ to $1.67$~nm, the two node positions change by only about $0.12\%$ and $0.82\%$, respectively, and $A_{\rm eff}^{(\rm node)}$ by $0.35\%$, whereas representative resonance heights change by $3$--$25\%$; details are collected in Appendix~\ref{app:convergence}.

Two further properties of the pair of zeros are worth recording.  First, the effective area is not a purely geometric quantity.  Repeating the node search for the conventional ring, where no splitting occurs, places the single zero at $B=0.248169$~T and gives $A_{\rm eff}/\pi r_0^2=1.06091$, about $2.6\%$ above the $\beta=0.30$ value.  Reading $A_{\rm eff}^{(s)}=\Phi_0/2B_s$ channel by channel,
\begin{equation}
A_{\rm eff}^{(\downarrow)}=1.03972\,\pi r_0^2,
\qquad
A_{\rm eff}^{(\uparrow)}=1.02802\,\pi r_0^2,
\label{eq:spin_areas}
\end{equation}
so the altermagnetic term makes the two spin species enclose effective areas differing by $1.13\%$.  The splitting is therefore most naturally read as a spin-dependent effective area, which also explains why $\Delta B$ converges more slowly than $\bar B$: it is the difference of two nearly equal quantities.  Measuring $A_{\rm eff}$ against a nonmagnetic reference ring of identical geometry would give a direct interferometric read-out of $\beta$.

Second, because the two zeros belong to different spin channels, the transmitted spin reverses completely between them.  Exactly at $B_\downarrow$ the $\downarrow$ channel vanishes and exactly at $B_\uparrow$ the $\uparrow$ channel does, so at either field the polarization saturates, $1-|P_T|<5\times10^{-9}$, limited only by the residual transmission left by the Green-function regularization; traversing the $2.9$~mT interval inverts the transmitted spin.  Near-unity purity survives slight detuning: the total-conductance minimum quoted above lies $3~\mu$T above $B_\downarrow$ and still gives $P_T=+0.998$.  The conductance stays below about $1~\mu$S across the interval, making this a high-purity and low-throughput reversal; it has no counterpart at $\beta=0$ and is a direct fingerprint of the two spin-resolved zeros.

\begin{figure*}[t]
\centering
\includegraphics[width=\textwidth]{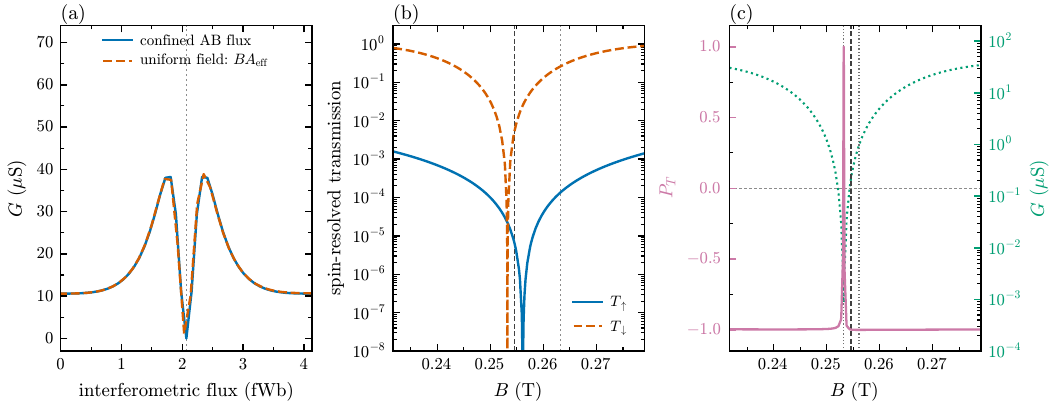}
\caption{Shift and spin resolution of the half-flux interference node for $\theta_A=\pi/8$. (a) Confined-flux conductance compared with the penetrating-field conductance plotted against the effective interferometric flux $BA_{\rm eff}$ rather than the nominal $B\pi r_0^2$.  The dominant field response is therefore a nearly rigid AB-phase translation.  (b) Spin-resolved transmissions near half flux.  The penetrating field splits the common zero into nearby nodes at $B_\downarrow$ and $B_\uparrow$; the gray dotted and black dashed guides mark the nominal and effective half-flux fields, respectively.  The field sampling is refined logarithmically at each zero, so both minima run below the plotted range; what the panel resolves is the separation of the two zeros, not their residual depths, which are set by the Green-function regularization.  (c) Corresponding spin polarization and total conductance.  The node positions are substantially more stable under mesh refinement than the heights of the neighboring coherent resonances.}
\label{fig:ab_vs_uniform}
\end{figure*}

\subsection{A flux-free high-throughput operating point}
\label{subsec:fluxfree}

The high-throughput point A introduced in Sec.~\ref{sec:combined} requires a confined flux of $0.689$~fWb threading a $50$~nm hole, which is experimentally demanding.  The uniform-field sweep shows that an equally good setting exists with no confined flux at all.  At
\begin{equation}
B=0.28961~\mathrm{T},\qquad \Phi_{\rm AB}=0,\qquad \theta_A=\pi/8,
\label{eq:pointF}
\end{equation}
the majority channel sits on a transmission maximum, $T_\downarrow=0.9999$, while the minority channel remains at $T_\uparrow=3.03\times10^{-3}$, so that $G=38.86~\mu$S and $P_T=-0.99395$.  Table~\ref{tab:points} compares this setting with points A and B.

\begin{table}[b]
\caption{Representative operating points at $\theta_A=\pi/8$ on the production grid.  Currents and current polarizations are at $V=0.254$~mV and $T=0$; $P_G$ is the polarization of the thermally averaged linear conductance.}
\label{tab:points}
\begin{ruledtabular}
\begin{tabular}{lccc}
 & A & B & flux-free\\
\hline
$B$ (T) & 0.13164 & 0.26328 & 0.28961\\
$\Phi_{\rm AB}$ (fWb) & 0.68928 & 0 & 0\\
$G$ ($\mu$S) & 38.84 & 10.20 & 38.86\\
$P_T$ & $-0.9919$ & $-0.9990$ & $-0.9940$\\
$I$ (nA) & 4.15 & 0.58 & 4.02\\
$P_I$ & $-0.959$ & $-0.988$ & $-0.964$\\
$P_G$ at $0.295$~K & $-0.980$ & $-0.991$ & $-0.982$\\
\end{tabular}
\end{ruledtabular}
\end{table}

The flux-free point matches the charge throughput of A, slightly improves both its transmission and its current polarization, and is marginally more robust to thermal averaging, while dispensing with the confined flux.  Its optimum is also broad: $G>37.8~\mu$S and $|P_T|>0.992$ are maintained over $0.2843\le B\le0.2944$~T, a $10$~mT window, in contrast with the millitesla scale of the interference zeros.  It is also at least as robust to static disorder: repeating the 24-realization Anderson ensembles of Sec.~\ref{sec:stage7} at this setting gives $\langle G\rangle=(37.98\pm0.66)~\mu$S with $\langle P_T\rangle=-0.9925\pm0.0018$ at $W=1.016$~meV, and $(35.72\pm2.20)~\mu$S with $-0.9871\pm0.0088$ at $W=2.032$~meV, every realization retaining $|P_T|>0.95$.  Of the settings examined here it is therefore the most accessible experimentally, since it requires only a uniform perpendicular field.

\subsection{Crystallographic switching and high-resistance state}

\begin{figure}[t]
\centering
\includegraphics[width=\columnwidth]{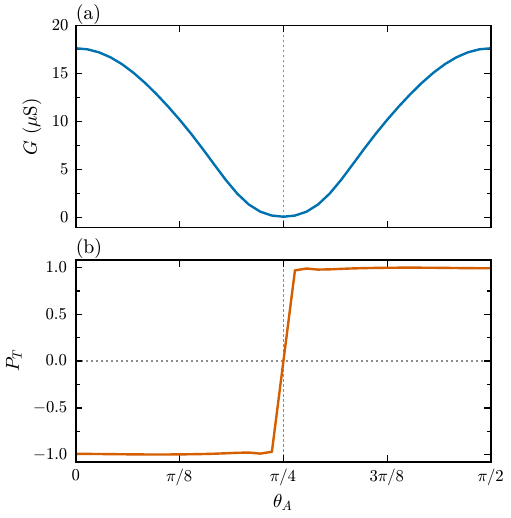}
\caption{Crystallographic control at $B=0.26328$~T. (a) Conductance over one complete $\pi/2$ period of the spin-summed response. (b) Transport spin polarization. A ninety-degree crystal rotation reverses the transmitted spin, while $\theta_A=\pi/4$ is exactly spin degenerate in the mirror-symmetric geometry.}
\label{fig:uniform_angular}
\end{figure}

At $B=0.26328$~T the crystal orientation acts as a strong transport control, as shown in Fig.~\ref{fig:uniform_angular}. A ninety-degree crystal rotation exchanges the two spin sectors and reverses $P_T$ without changing the total conductance. Representative values are
\begin{align}
\theta_A=0:&\quad G\simeq18~\mu\mathrm{S},\quad P_T=-0.994,\\
\theta_A=\pi/8:&\quad G\simeq10~\mu\mathrm{S},\quad P_T=-0.999,\\
\theta_A=\pi/4:&\quad G\simeq0.11~\mu\mathrm{S},\quad P_T=0,\\
\theta_A=3\pi/8:&\quad G\simeq10~\mu\mathrm{S},\quad P_T=+0.999.
\label{eq:uniform_angular_values}
\end{align}
The mirror orientation therefore realizes a deep, spin-degenerate interference minimum.  Because $R=1/G$ magnifies small grid-dependent changes near a transmission zero, we do not assign a bulk magnetoresistance coefficient to this state; the robust statement is the crystallographically controlled suppression of conductance together with exact $P_T=0$.

\subsection{Energy--field structure}

\begin{figure*}[t]
\centering
\includegraphics[width=6.632in]{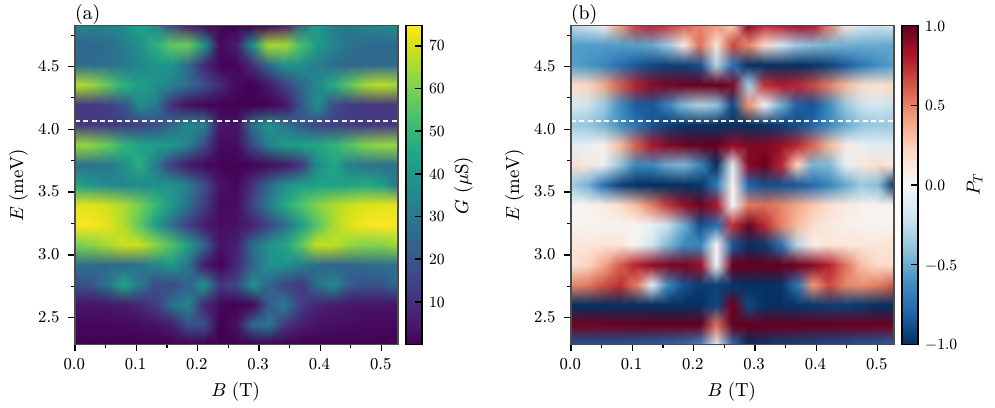}
\caption{Uniform-field transport maps at $\theta_A=\pi/8$. (a) Conductance and (b) spin polarization as functions of carrier energy and magnetic field. The physical scales correspond to the $50$~nm reference ring and $m^*=0.30m_e$.}
\label{fig:uniform_map}
\end{figure*}

The energy--field maps in Fig.~\ref{fig:uniform_map} show that the spin-selective response is not confined to a single isolated point. Resonant structures move with field predominantly because the interference phase is displaced by $BA_{\rm eff}$, with smaller finite-width orbital corrections, and extended regions with large $|P_T|$ coexist with measurable conductance. These maps are used to identify operating windows; quantitative conclusions are based on the refined one-dimensional cuts discussed above.

\section{Combined orbital field and confined Aharonov--Bohm flux}
\label{sec:combined}

We now activate both orbital controls. The uniform-field Wilson line and the singular-gauge AB phase multiply on every central-region hopping, while only the uniform field enters the lead self-energies. The confined flux therefore controls the winding phase independently of the orbital field within the conducting material.

\subsection{Combined-field symmetry constraints}

Flux-quantum periodicity remains exact,
\begin{equation}
G(B,\Phi_{\rm AB}+\Phi_0,\theta_A)=G(B,\Phi_{\rm AB},\theta_A),
\label{eq:combined_periodicity}
\end{equation}
whereas at fixed nonzero $B$ the separate symmetry $\Phi_{\rm AB}\rightarrow-\Phi_{\rm AB}$ is no longer required. Generalized two-terminal reciprocity \cite{Buttiker1986} instead gives the simultaneous reversal
\begin{equation}
G(B,\Phi_{\rm AB},\theta_A)=G(-B,-\Phi_{\rm AB},\theta_A).
\label{eq:combined_reciprocity}
\end{equation}
The angular spin-exchange relation under $\theta_A\rightarrow\theta_A+\pi/2$ remains intact. At the mirror orientation $\theta_A=\pi/4$, Eqs.~\eqref{eq:mirror_relation} and \eqref{eq:pi4_degeneracy} guarantee exact spin degeneracy even when both orbital controls are present.

\subsection{Field-tunable AB phase}

\begin{figure*}[t]
\centering
\includegraphics[width=\textwidth]{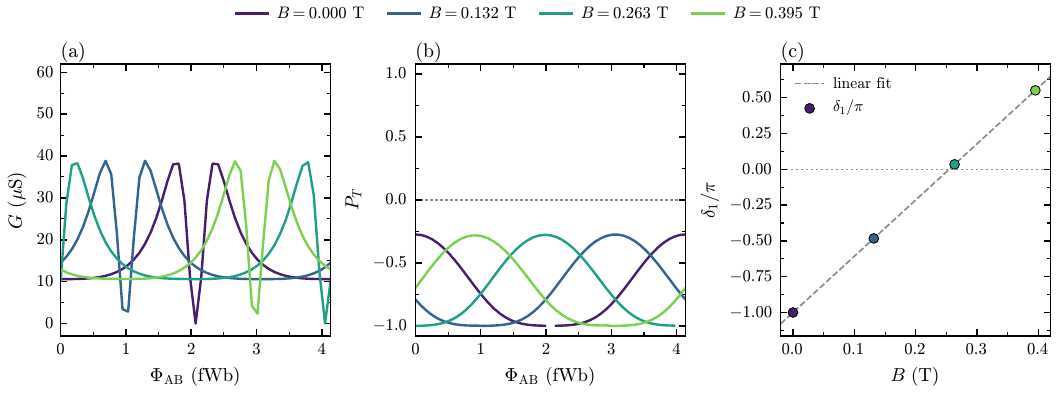}
\caption{Combined uniform-field and confined-flux transport at $E_F=4.064$~meV, $\beta=0.30$, and $\theta_A=\pi/8$. (a) Conductance over one AB period for selected magnetic fields. (b) Spin polarization; points with conductance below $1.94~\mu$S are omitted from the ratio plot. (c) Phase of the first AB harmonic extracted from the same cuts. The phase varies linearly with $B$; its slope yields $A_{\rm eff}=1.0342\,\pi r_0^2$, while the confined-flux period remains $\Phi_0=4.13567$~fWb.}
\label{fig:stage5_phase}
\end{figure*}

Figure~\ref{fig:stage5_phase} shows confined-flux sweeps at $B=0$, $0.132$, $0.263$, and $0.395$~T. At zero uniform field the curve is even in $\Phi_{\rm AB}$. Once $B$ is finite, the left and right parts of a flux period are no longer constrained to coincide. The dense cuts show that the corresponding asymmetry is of comparable magnitude at all three finite fields, consistent with the nearly rigid phase translation rather than a separate field-dependent amplitude mechanism. The first harmonic
\begin{equation}
G(\Phi_{\rm AB})=G_0+A_1\cos\!\left(2\pi\frac{\Phi_{\rm AB}}{\Phi_0}-\delta_1\right)+\cdots
\label{eq:first_harmonic}
\end{equation}
acquires a continuously field-dependent phase. Using 48 confined-flux points per period to avoid harmonic aliasing, the unwrapped fitted phases are $\delta_1/\pi=-1.0000$, $-0.4828$, $0.0343$, and $0.5513$ at the four fields above.  The corresponding first-harmonic amplitudes are $A_1=9.33$, $9.33$, $9.34$, and $9.36$~$\mu$S, while the mean conductance changes only from $G_0=17.56$ to $17.62$~$\mu$S (less than $0.4\%$).  Thus the dominant action of $B$ is a nearly rigid phase translation,
\begin{equation}
\delta_1(B)\simeq\delta_1(0)+2\pi\frac{BA_{\rm eff}}{\Phi_0},
\label{eq:phase_area_relation}
\end{equation}
with $A_{\rm eff}$ given by Eq.~\eqref{eq:Aeff}.  After compensating this phase shift, the mean absolute residual between the finite-$B$ and shifted $B=0$ conductance curves is only $0.26$--$0.39~\mu$S over the sampled fields.  This residual is the direct measure of finite-width orbital reconstruction. The dense Fourier extraction and the effective-area fit are documented in Appendix~\ref{app:dense_fourier}.

\subsection{Dual control of spin-polarized transmission}

The confined flux also provides an independent control of the spin imbalance created by the orbital field. At $\theta_A=\pi/8$ and $B=0.26328$~T,
\begin{align}
&\Phi_{\rm AB}=0:\quad G\simeq10~\mu\mathrm{S},\quad P_T=-0.99895,\\
&\Phi_{\rm AB}=2.06783~\mathrm{fWb}:\quad G\simeq11~\mu\mathrm{S},\quad P_T=-0.280.
\label{eq:AB_gate_example}
\end{align}
Thus a half-quantum change of the confined flux strongly suppresses spin polarization while leaving the charge conductance of the same order. Conversely, the high-throughput operating point
\begin{equation}
B=0.13164~\mathrm{T},\quad
\Phi_{\rm AB}=0.68928~\mathrm{fWb},\quad
\theta_A=\pi/8
\label{eq:pointA}
\end{equation}
gives
\begin{equation}
G\simeq39~\mu\mathrm{S},\qquad P_T=-0.9919.
\label{eq:highG_filter}
\end{equation}

\begin{figure*}[t]
\centering
\includegraphics[width=6.632in]{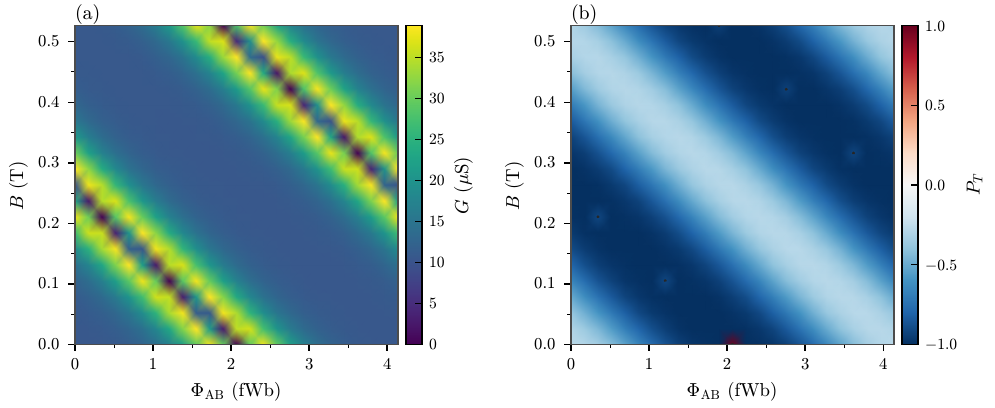}
\caption{Two-parameter control map at $\theta_A=\pi/8$. (a) Conductance and (b) spin polarization versus uniform magnetic field and confined AB flux. Polarization is masked where $G<1.94~\mu$S. The map is used to identify operating regions; the refined one-dimensional cuts provide the quantitative values quoted in the text.}
\label{fig:stage5_maps}
\end{figure*}

The two-parameter map in Fig.~\ref{fig:stage5_maps} surveys $0\le B\le0.5266$~T and $0\le\Phi_{\rm AB}\le4.1357$~fWb. Extended regions combine measurable conductance and $|P_T|>0.9$, demonstrating that the spin-filter response is not restricted to a single isolated resonance.

\begin{figure*}[t]
\centering
\includegraphics[width=5.644in]{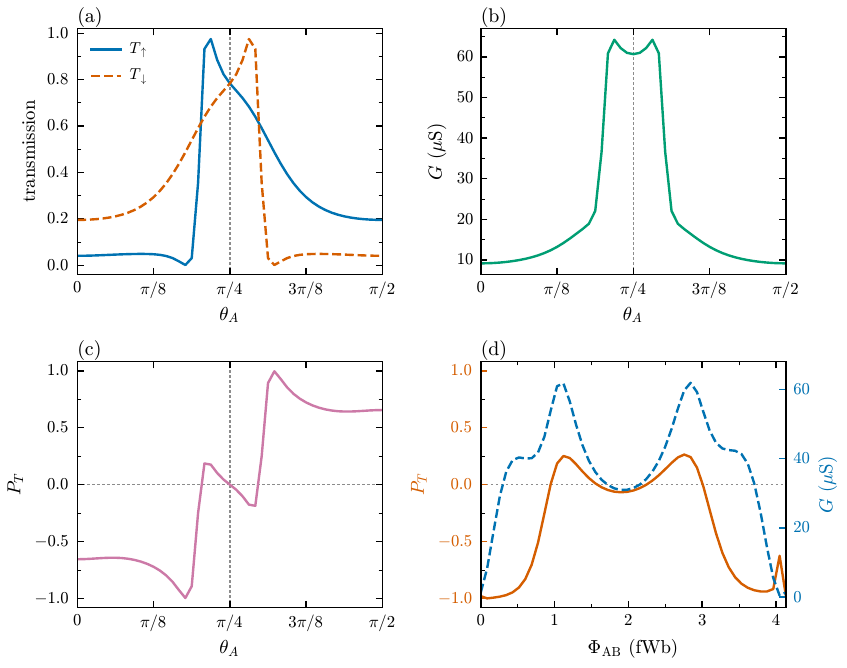}
\caption{Combined crystallographic and confined-flux control. Panels (a)--(c) show spin transmissions, total conductance, and polarization versus $\theta_A$ at $B=0.26328$~T and $\Phi_{\rm AB}=1.03392$~fWb. The mirror orientation $\theta_A=\pi/4$ remains spin degenerate. Panel (d) shows a flux-driven sign change of $P_T$ at fixed $\theta_A=5\pi/24$ and $B=0.26328$~T; the dashed curve is the corresponding conductance.}
\label{fig:stage5_switching}
\end{figure*}

Crystallographic orientation supplies a third independent control. At $B=0.26328$~T and $\Phi_{\rm AB}=1.03392$~fWb, the refined angular scan in Fig.~\ref{fig:stage5_switching} passes through the exact spin-degenerate point $\theta_A=\pi/4$ and reverses the sign of $P_T$ across mirror-related orientations. At fixed $\theta_A=5\pi/24$ and the same field, a confined-flux sweep can also change the sign of $P_T$ while retaining finite conductance, although this flux-driven inversion is more resonance sensitive than the crystallographic spin switch.

\section{Finite-bias transport and thermal broadening}
\label{sec:finitebias_results}

The linear-response operating points are useful only if their spin selectivity survives integration over a finite energy window. We evaluate Eq.~\eqref{eq:current} with symmetric electrochemical potentials
\begin{equation}
\mu_{L,R}=E_F\pm\frac{eV}{2}.
\label{eq:symmetric_bias}
\end{equation}
The scattering Hamiltonian is kept bias independent, so this is a frozen-scatterer Landauer treatment appropriate to modest voltages. At zero temperature,
\begin{equation}
I_s(V)=\frac{e}{h}\int_{E_F-e|V|/2}^{E_F+e|V|/2}\!dE\,T_s(E)\,\operatorname{sgn}(V).
\label{eq:physical_current}
\end{equation}

\subsection{Energy-resolved spin-filter windows}

\begin{figure*}[t]
\centering
\includegraphics[width=6.632in]{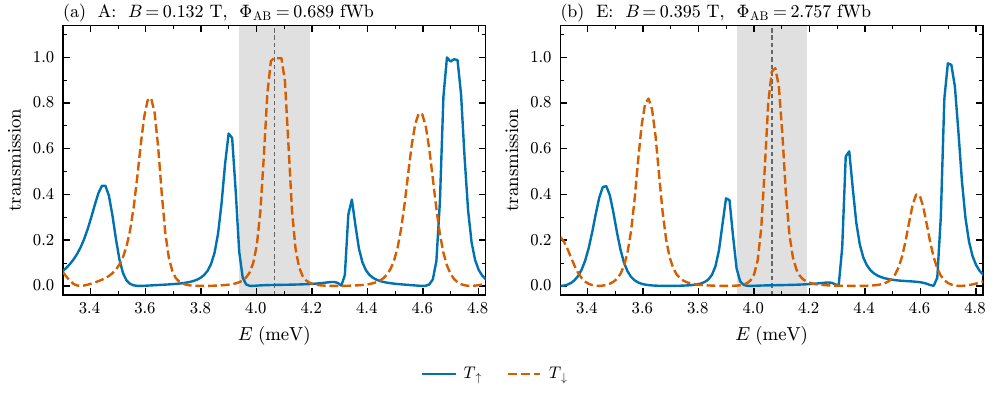}
\caption{Energy-resolved spin transmission. (a) High-throughput point A, $B=0.132$~T and $\Phi_{\rm AB}=0.689$~fWb. (b) Broader-window point E, $B=0.395$~T and $\Phi_{\rm AB}=2.757$~fWb. The dashed line marks $E_F=4.064$~meV and the shaded region is the symmetric zero-temperature transport window for $|V|=0.254$~mV.}
\label{fig:stage6_energy}
\end{figure*}

Figure~\ref{fig:stage6_energy} resolves the transmissions from $3.302$ to $4.826$~meV. Point A is the high-throughput filter of Eq.~\eqref{eq:pointA}. A second operating point,
\begin{equation}
B=0.39493~\mathrm{T},\quad
\Phi_{\rm AB}=2.75711~\mathrm{fWb},\quad
\theta_A=\pi/8,
\label{eq:pointE}
\end{equation}
has representative-grid $G(E_F)\simeq36~\mu$S and $P_T(E_F)=-0.9944$. The dominant spin channel forms a narrow resonance around $E_F=4.064$~meV, with opposite-spin resonances nearby. The shaded interval in Fig.~\ref{fig:stage6_energy} is the zero-temperature transport window for $|V|=0.254$~mV.

\subsection{Current--voltage characteristics}

\begin{figure*}[t]
\centering
\includegraphics[width=6.632in]{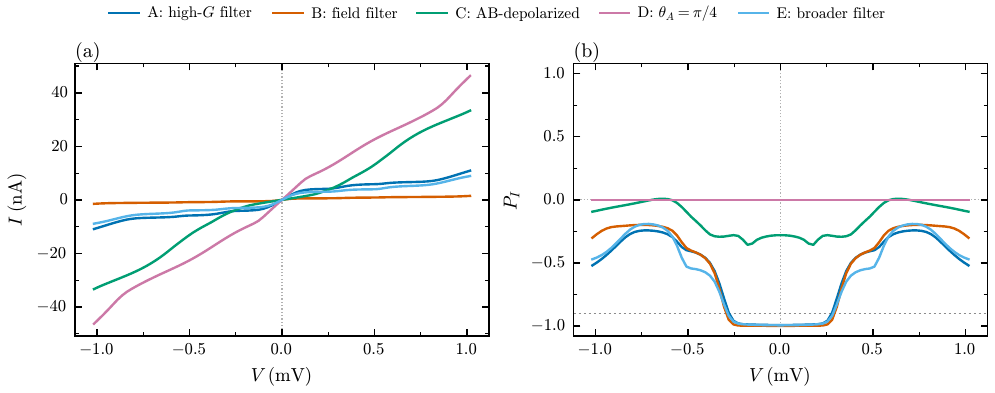}
\caption{Finite-bias coherent transport at zero temperature. (a) Total current in nA. (b) Current polarization. A and E are high-throughput/high-polarization settings, B is the low-throughput field filter, C is the AB-depolarized control, and D is the mirror-symmetric control. The dotted line marks $P_I=-0.9$.}
\label{fig:stage6_IV}
\end{figure*}

Figure~\ref{fig:stage6_IV} shows the zero-temperature current and current polarization. At point A,
\begin{align}
V=0.254~\mathrm{mV}:&\quad I\simeq4.2~\mathrm{nA},\quad P_I=-0.959,\\
V=0.508~\mathrm{mV}:&\quad I\simeq5.9~\mathrm{nA},\quad P_I=-0.400.
\label{eq:pointA_IV}
\end{align}
The near-perfect spin-filter window is therefore finite rather than broadband. Point E gives $I\simeq3.1$~nA and $P_I=-0.965$ at $0.254$~mV. The field-dominated operating point B, $B=0.26328$~T and $\Phi_{\rm AB}=0$, sits on the high-field flank of the $T_\uparrow$ zero located in Sec.~\ref{sec:uniformB}, $7.1$~mT above it and at roughly a quarter of the local conductance maximum.  Its higher purity ($P_I=-0.988$) and its small current ($0.58$~nA) are two aspects of the same fact: suppressing the minority channel by approaching its interference zero suppresses the charge throughput as well.  Points A, E, and the flux-free setting of Eq.~\eqref{eq:pointF} operate the other way round, placing the majority channel on a transmission maximum, and therefore combine high polarization with a current roughly seven times larger.  Node proximity and majority resonance are thus the two ends of the purity-throughput trade-off available in this device.

Two controls clarify the interpretation. At point C, $B=0.26328$~T, $\Phi_{\rm AB}=2.06783$~fWb, and $\theta_A=\pi/8$, one finds $I\simeq3.9$~nA but only $P_I=-0.280$ at $0.254$~mV. At the mirror point D, $B=0.26328$~T, $\Phi_{\rm AB}=1.03392$~fWb, and $\theta_A=\pi/4$, $T_\uparrow(E)=T_\downarrow(E)$ throughout the sampled energy range; hence $P_I=0$ while the charge current reaches about $12$~nA at $0.254$~mV.

\subsection{Thermal broadening}

Finite temperature introduces an independent energy average. We define
\begin{equation}
P_G(T)=\frac{G_\uparrow(T)-G_\downarrow(T)}{G_\uparrow(T)+G_\downarrow(T)},
\label{eq:PG}
\end{equation}
with $G_s(T)$ obtained from Eq.~\eqref{eq:finiteT}. At $T=0.295$~K, the linear-response polarizations are $-0.980$, $-0.991$, and $-0.985$ for A, B, and E, respectively. At the same temperature and $V=0.254$~mV, the corresponding current polarizations decrease to $-0.829$, $-0.854$, and $-0.867$. At $T=0.147$~K, point A still gives $P_I=-0.904$ at $0.254$~mV, while B and E give $-0.937$ and $-0.924$.

\begin{figure*}[t]
\centering
\includegraphics[width=6.349in]{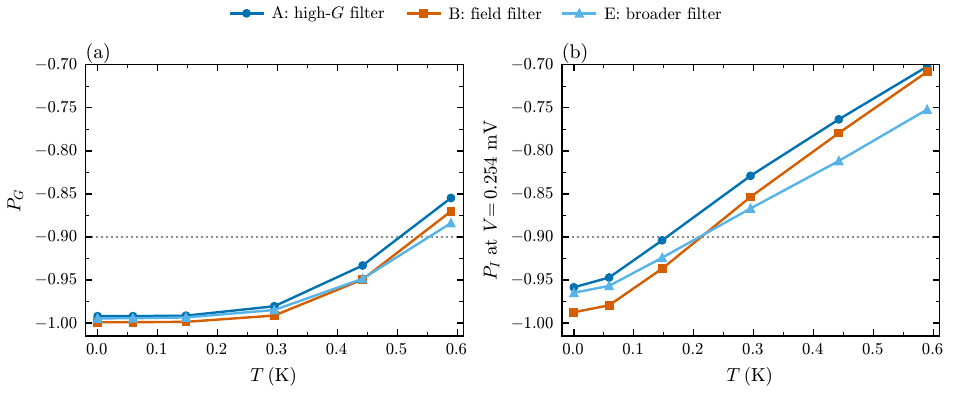}
\caption{Thermal broadening of the spin-filter settings. (a) Polarization of the thermally averaged linear conductance. (b) Finite-bias current polarization at $V=0.254$~mV. The dotted line marks $-0.9$. The useful regime is a sub-kelvin, low-bias window for the chosen reference mass and geometry.}
\label{fig:stage6_thermal}
\end{figure*}

The strongest filters are therefore resonance windows rather than broad spin gaps. This is physically useful but sets a clear experimental requirement: the voltage and thermal energy windows must remain smaller than the separation to nearby opposite-spin resonances. Quantitative extension to substantially larger voltages requires a self-consistent electrostatic potential.

\subsection{Zeeman-scale estimate}

The baseline model intentionally omits a Zeeman term in order to isolate the nonrelativistic altermagnetic splitting.  For the sub-tesla fields used here, however, the scale is not identically negligible.  Taking $g=2$ gives $g\mu_BB=0.0152$, $0.0305$, and $0.0457$~meV at the representative fields $B=0.132$, $0.263$, and $0.395$~T.  If a spatially uniform $g$ factor acts throughout the central region and ideal leads and $\sigma_z$ remains conserved, the Zeeman contribution is incorporated exactly as an opposite rigid energy shift of the two spin blocks, $T_s^{(Z)}(E_F)=T_s^{(0)}(E_F-sg\mu_BB/2)$.  Direct NEGF evaluation at the shifted energies gives $P_T=-0.9922$, $-0.9985$, and $-0.9956$ at points A, B, and E, respectively.  The resonance amplitudes can change appreciably (most strongly at point B), but the near-unity spin selectivity survives.  The persistence of the polarization under this shift shows that the filtering is not an ordinary Zeeman effect; it comes from the altermagnetic interference pattern.  Detailed numbers are listed in Appendix~\ref{app:zeeman}.

\section{Static disorder and phase relaxation}
\label{sec:stage7}

We next test whether the high-throughput point A survives static scalar disorder and phase relaxation.  The disorder calculation is read in the usual mesoscopic sense, where sample-specific resonances fluctuate even when symmetry-protected trends remain stable \cite{Anderson1958,LeeStone1985,LeeStoneFukuyama1987,Beenakker1997}. In physical units this point is
\begin{equation}
B=0.13164~\mathrm{T},\quad
\Phi_{\rm AB}=0.68928~\mathrm{fWb},\quad
\theta_A=\pi/8,
\label{eq:stage7_pointA}
\end{equation}
with representative-grid $G\simeq39~\mu$S and $P_T=-0.9919$.

\subsection{Nonmagnetic onsite disorder}

We add a spin-independent Anderson term in the conducting annulus,
\begin{equation}
H_{\rm dis}=\sum_i \xi_i c_i^\dagger c_i,
\qquad
\xi_i\in[-W/2,W/2],
\label{eq:anderson_disorder}
\end{equation}
using the same realization in the two conserved spin blocks. We average over 24 independent realizations for each disorder width between $0.254$ and $4.064$~meV at points A and E. At point A, $W=1.016$~meV gives
\begin{equation}
\langle G\rangle=(38.03\pm0.64)~\mu\mathrm{S},\qquad
\langle P_T\rangle=-0.9903\pm0.0017,
\end{equation}
where the uncertainties are sample standard deviations. At $W=2.032$~meV,
\begin{equation}
\langle G\rangle=(35.90\pm2.15)~\mu\mathrm{S},\qquad
\langle P_T\rangle=-0.9846\pm0.0082.
\label{eq:disorderA_physical}
\end{equation}
Point E behaves similarly, with $(32.72\pm4.10)~\mu$S and $-0.9859\pm0.0099$ at $W=2.032$~meV. All 48 realizations at this disorder strength retain $|P_T|>0.95$. Even at $W=4.064$~meV the ensemble means remain $|\langle P_T\rangle|\simeq0.955$ at A and $0.952$ at E, although the conductance fluctuations become broad. Thus nonmagnetic disorder degrades charge throughput more rapidly than spin selectivity in the operating windows considered here.

\subsection{Spin-conserving B\"uttiker probes}

\begin{figure}[!htb]
\centering
\includegraphics[width=\columnwidth]{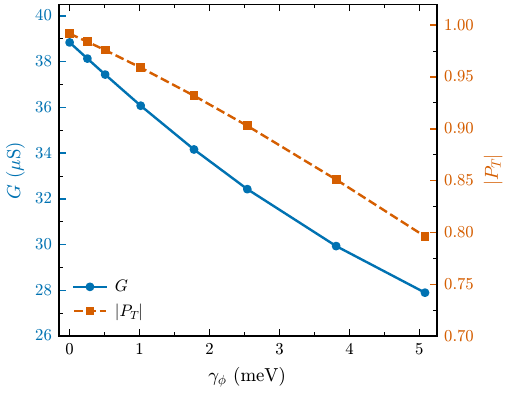}
\caption{Coherence-to-incoherence crossover of operating point A. Spin-conserving B\"uttiker voltage probes reduce the conductance and transmitted spin polarization smoothly as the phenomenological probe coupling $\gamma_\phi$ increases.}
\label{fig:stage7_dephasing}
\end{figure}

To model phase relaxation without spin flips, twelve wide-band B\"uttiker voltage probes are attached near the mean radius. For each spin block,
\begin{equation}
\Sigma^r_{p,s}=-\frac{\ii\gamma_\phi}{2}|p\rangle\langle p|,
\qquad
\Gamma_{p,s}=\gamma_\phi|p\rangle\langle p|,
\label{eq:probe_self_energy}
\end{equation}
and the probe electrochemical potentials are fixed by $I_{p,s}=0$ \cite{ButtikerPRB1986Coherence,DAmatoPastawski1990,Forster2007}. The zero-current constraints are solved independently for the two conserved spin sectors.

Figure~\ref{fig:stage7_dephasing} shows a gradual coherence crossover. At $\gamma_\phi=1.016$~meV,
\begin{equation}
G=36.07~\mu\mathrm{S},\qquad |P_T|=0.9591,
\end{equation}
while at $\gamma_\phi=2.540$~meV,
\begin{equation}
G=32.43~\mu\mathrm{S},\qquad |P_T|=0.9029.
\end{equation}
Even at $\gamma_\phi=5.080$~meV the polarization remains $0.796$. The parameter $\gamma_\phi$ is a phenomenological energy broadening; associating it with a unique phase-coherence length would require a microscopic scattering model and a material-specific group velocity.

The filtering does not rely on perfect coherence. Weak scalar disorder mainly changes the transmitted charge, whereas stronger coupling to the spin-conserving voltage probes gradually repopulates the minority channel.

A self-consistent finite-bias Poisson--NEGF calculation is not introduced because it requires a specific dielectric environment, gate geometry, electrostatic boundary conditions, and contact work-function alignment. The finite-bias results above should be interpreted as the frozen-scatterer limit until such a device electrostatics is specified.

\section{Discussion}
\label{sec:discussion}

Three controls play different roles in the device.  The crystallographic angle fixes the orientation of the $d$-wave altermagnetic tensor relative to the contacts. A confined AB flux changes only the winding phase and preserves exact $h/e$ periodicity. A uniform field penetrates the annulus and the leads, but over the field range studied its dominant interferometric effect is quantitatively captured by the effective area $A_{\rm eff}=1.0342\,\pi r_0^2$: it translates the AB phase nearly rigidly. After compensating this phase shift, the mean absolute residual remains below $0.4~\mu$S for each sampled cut, while localized deviations near narrow resonances can reach a few $\mu$S; these residuals encode finite-width orbital reconstruction. When both controls coexist, the confined-flux period remains $4.13567$~fWb and the first-harmonic phase becomes continuously field tunable.

For the $50$~nm reference ring the relevant magnetic fields are modest: the most important operating points occur between about $0.13$ and $0.40$~T. The half-flux comparison is especially transparent. A confined flux of $2.068$~fWb produces the exact one-mode node, while a penetrating field shifts that cancellation to the effective-area scale near $0.255$~T and resolves it into two nearby spin-dependent zeros. Their mean position yields $A_{\rm eff}^{(\rm node)}=1.03384\,\pi r_0^2$, reproducing the Fourier-phase area to $0.04\%$. The penetrating field therefore shifts and spin-resolves the common AB node rather than simply lifting it at the nominal mean-area half-flux field.  On the flank of the $T_\uparrow$ zero, for example at $B=0.263$~T, a strongly polarized channel with $|P_T|>0.998$ remains available without ferromagnetic leads, while placing the majority channel on resonance at $B=0.290$~T gives comparable purity with nearly four times the conductance and no confined flux. The explicit $g=2$ estimate above shows that adding an ordinary Zeeman scale changes resonance amplitudes but does not account for the near-unity polarization.

Crystallographic rotation supplies an independent switch. At $B=0.263$~T, moving from $\theta_A=\pi/8$ to $3\pi/8$ reverses the transmitted spin while preserving the total conductance. The intermediate mirror orientation is unpolarized and highly resistive. This angular structure is more robust than the absolute height of individual coherent resonances and is the central device signature of the altermagnetic tensor.

The finite-bias and finite-temperature results quantify the experimental coherence window. Point A carries about $4.2$~nA with $P_I=-0.959$ at $0.254$~mV and $T=0$, but nearby opposite-spin resonances dilute the polarization as the voltage increases. Thermal broadening similarly places the chosen reference device in a sub-kelvin operating regime. These temperature and voltage scales are not fundamental material limits: they scale with $\hbar^2/(2m^*\ell^2)$ and can be shifted by changing the ring dimensions or the effective mass.

The isolated circular ring retains the $\Delta m=\pm2$ selection rule of the $d$-wave operator and associated anticrossings.  In the open device $m$ is not conserved, so a quantitative one-to-one assignment of every transport resonance to a closed-ring anticrossing would require a dedicated pole/local-density-of-states analysis.  We therefore regard that spectroscopy as an outlook rather than as a result claimed here.

The connection to $\mathrm{Rb}_{1-\delta}\mathrm{V}_2\mathrm{Te}_2\mathrm{O}$ should be interpreted carefully. The experimentally observed layered metallic state and crystal-symmetry-paired spin splitting make it a compelling platform for a lithographically defined interferometer \cite{ZhangNatPhys2025RbV,HuPRX2025RbV,MaoPRB2026RbV}. At the same time, recent neutron diffraction finds a G-type antiferromagnetic structure below $337$~K \cite{XieCPB2026RbV}. The present model therefore does not claim a quantitative first-principles description of that measured magnetic structure. Rather, it supplies a symmetry-controlled mesoscopic effective model and an SI-scale device proposal that can later be refitted to a selected experimental pocket once its transport mass, domain structure, and interface electrostatics are known.

\section{Conclusions}
\label{sec:conclusions}

For the finite-width $d$-wave altermagnetic ring studied here, the clearest transport signatures are the crystal-angle spin switch, the exact $h/e$ AB periodicity, and the interference zeros.  A perpendicular field translates the AB pattern with an effective area $A_{\rm eff}=1.0342\,\pi r_0^2$ and splits the common half-flux zero into two nearby spin-resolved nodes.  Their mean position gives $A_{\rm eff}^{(\rm node)}=1.0338\,\pi r_0^2$, in close agreement with the phase-slope determination.  These node positions, the symmetry relations, and the spin polarization are numerically more stable than the heights of individual coherent resonances.  In the $50$~nm reference device, sub-tesla fields produce windows with $|P_T|>0.99$ and nA-scale polarized currents at sub-mV bias.  The most accessible of these needs no confined flux at all: a uniform field of $0.290$~T alone places the majority channel on a transmission maximum and yields $G=38.9~\mu$S with $P_T=-0.994$ over a $10$~mT window.

The same spin selectivity persists when the ideal coherent picture is relaxed.  For a $2.03$~meV Anderson disorder width, 24-realization ensembles at two operating points retain mean $|P_T|\simeq0.985$.  Coupling to spin-conserving B\"uttiker voltage probes is the stronger depolarizing mechanism, but it acts gradually, leaving $|P_T|\simeq0.90$ at a probe broadening of $2.54$~meV.  The device should therefore be viewed as a tunable mesoscopic interferometer rather than as a broadband perfect filter: crystal orientation, confined flux, penetrating field, voltage, temperature, disorder, and phase relaxation all control the transmitted spin in distinct ways.

$\mathrm{Rb}_{1-\delta}\mathrm{V}_2\mathrm{Te}_2\mathrm{O}$ supplies a realistic materials scale and symmetry motivation, but a quantitative prediction for a fabricated structure will require an experimentally constrained low-energy mass tensor, domain configuration, and self-consistent electrostatics.  Those inputs are needed before moving from the present effective-device model to a material-specific Poisson--NEGF calculation.

\appendix

\section{Rotated form of the \texorpdfstring{$d$}{d}-wave altermagnetic operator}
\label{app:rotation}

Let $(x',y')$ denote crystal axes obtained by rotating the laboratory axes by $\theta_A$. The kinetic momenta transform as
\begin{align}
\Pi_{x'}&=\Pi_x\cos\theta_A+\Pi_y\sin\theta_A,
\label{eq:pixprime}\\
\Pi_{y'}&=-\Pi_x\sin\theta_A+\Pi_y\cos\theta_A.
\label{eq:piyprime}
\end{align}
Expanding the difference of squares gives
\begin{align}
\Pi_{x'}^2-\Pi_{y'}^2
={}&(\cos^2\theta_A-\sin^2\theta_A)(\Pi_x^2-\Pi_y^2)
\nonumber\\
&+2\sin\theta_A\cos\theta_A\{\Pi_x,\Pi_y\}
\nonumber\\
={}&\cos(2\theta_A)(\Pi_x^2-\Pi_y^2)
+\sin(2\theta_A)\{\Pi_x,\Pi_y\},
\label{eq:rotated_operator}
\end{align}
which yields Eq.~\eqref{eq:HAM}. The anticommutator is required because $\Pi_x$ and $\Pi_y$ do not commute in a perpendicular magnetic field.

Introducing $\Pi_\pm=\Pi_x\pm\ii\Pi_y$, one may also write
\begin{equation}
\Pi_{x'}^2-\Pi_{y'}^2
=\frac{1}{2}\left(
 e^{-2\ii\theta_A}\Pi_+^2
+e^{2\ii\theta_A}\Pi_-^2
\right),
\label{eq:ladder_form}
\end{equation}
which makes the $\Delta m=\pm2$ structure of the circular closed-ring problem explicit.

\section{Finite-difference representation of the mixed derivative}
\label{app:stencil}

At zero vector potential,
\begin{equation}
\{\Pi_x,\Pi_y\}=-2\hbar^2\partial_x\partial_y.
\label{eq:mixed_zeroB}
\end{equation}
For a square finite-difference spacing $\Delta$,
\begin{align}
\partial_x\partial_y\psi_{i,j}
\simeq\frac{1}{4\Delta^2}\Big(&
\psi_{i+1,j+1}-\psi_{i+1,j-1}
\nonumber\\
&-\psi_{i-1,j+1}+\psi_{i-1,j-1}
\Big),
\label{eq:central_mixed}
\end{align}
with truncation error of order $\Delta^2$. Using the physical kinetic hopping energy $t_\Delta$ defined in Eq.~\eqref{eq:tDelta}, the mixed altermagnetic term generates diagonal hoppings of magnitude
\begin{equation}
|t_{xy,s}|=\frac{|s\beta\sin2\theta_A|}{2}\,t_\Delta,
\label{eq:txy}
\end{equation}
with alternating signs on the two lattice diagonals, as used in Eqs.~\eqref{eq:hpp} and \eqref{eq:hpm}.

\section{Finite-difference convergence of representative operating points}
\label{app:convergence}

The production figures use a square finite-difference spacing $\Delta=2.14$~nm. Because the coherent conductance contains narrow resonances, their heights converge substantially more slowly than the spin polarization and the positions of interference zeros. Table~\ref{tab:conv} reports the same physical operating points while refining $\Delta$ at fixed physical $E_F$, $B$, $\Phi_{\rm AB}$, ring geometry, effective mass, and a strictly fixed $30$~nm contact aperture.  The allowed refinement factors satisfy $(W_{\rm lead}+1)\Delta=30$~nm exactly; the code rejects grids that would silently change the physical opening when enforcing an odd lead-site count. The absolute conductance can move substantially as a narrow resonance shifts relative to $E_F$, whereas the sign and near-unity magnitude of $P_T$ are much more stable. This is why the main text treats $G$, $I$, and $R$ as representative device values and bases the central conclusions on symmetry, periodicity, phase, node positions, and polarization.

\begin{table*}[t]
\caption{Mesh refinement at points A, B, and E. Conductance is in $\mu$S.}
\label{tab:conv}
\begin{ruledtabular}
\begin{tabular}{ccccccc}
$\Delta$ (nm) & $G_A$ & $P_A$ & $G_B$ & $P_B$ & $G_E$ & $P_E$\\
\hline
2.14 & 38.84 & $-0.9919$ & 10.20 & $-0.9990$ & 36.44 & $-0.9944$\\
1.88 & 38.52 & $-0.9941$ & 12.13 & $-0.9992$ & 36.32 & $-0.9962$\\
1.67 & 37.58 & $-0.9924$ & 12.76 & $-0.9996$ & 38.29 & $-0.9945$\\
1.50 & 36.23 & $-0.9929$ & 14.26 & $-0.9996$ & 38.76 & $-0.9953$\\
1.36 & 33.76 & $-0.9943$ & 19.59 & $-0.9999$ & 38.79 & $-0.9962$\\
1.25 & 31.89 & $-0.9928$ & 18.86 & $-0.9999$ & 38.05 & $-0.9954$\\
1.15 & 29.32 & $-0.9924$ & 21.89 & $-1.0000$ & 36.82 & $-0.9949$\\
1.07 & 28.24 & $-0.9944$ & 23.94 & $-1.0000$ & 37.42 & $-0.9963$\\
1.00 & 24.90 & $-0.9922$ & 28.21 & $-1.0000$ & 34.14 & $-0.9950$\\
\end{tabular}
\end{ruledtabular}
\end{table*}

The nonmonotonic conductance convergence is characteristic of resolving a narrow coherent resonance at a fixed Fermi energy; it does not support a reliable Richardson extrapolation of the resonance height. By contrast, $|P_T|\ge0.9918$ throughout the fixed-aperture refinement sequence.  The node positions provide an additional hierarchy check: between $\Delta=2.14$ and $1.67$~nm, $B_\downarrow$ and $B_\uparrow$ shift by approximately $0.12\%$ and $0.82\%$, respectively, and $A_{\rm eff}^{(\rm node)}/(\pi r_0^2)$ changes from $1.03384$ to $1.03022$ ($0.35\%$), while representative resonance heights change by about $3$--$25\%$.  The splitting itself changes from $2.88$ to $5.30$~mT and is therefore reported only as being of order a few millitesla.  Table~\ref{tab:nodes} collects these numbers.  A material-specific prediction of absolute current would require both a better constrained low-energy band model and a physical broadening mechanism rather than the ideal coherent-resonance limit.

\begin{table}[!htb]
\caption{Spin-resolved half-flux zeros at $\theta_A=\pi/8$ and $\Phi_{\rm AB}=0$, located by bounded minimization of each spin-resolved transmission.  Both mesh spacings share the same fixed $30$~nm contact aperture.}
\label{tab:nodes}
\begin{ruledtabular}
\begin{tabular}{ccccc}
$\Delta$ (nm) & $B_\downarrow$ (T) & $B_\uparrow$ (T) & $\Delta B$ (mT) & $A_{\rm eff}^{(\rm node)}/\pi r_0^2$\\
\hline
2.14 & 0.253226 & 0.256108 & 2.88 & 1.03384\\
1.67 & 0.252911 & 0.258213 & 5.30 & 1.03022\\
\hline
change & $0.12\%$ & $0.82\%$ & $84\%$ & $0.35\%$\\
\end{tabular}
\end{ruledtabular}
\end{table}

\section{Zeeman-scale estimate}
\label{app:zeeman}

The baseline Hamiltonian omits Zeeman coupling in order to isolate the altermagnetic mechanism. For a spatially uniform $g=2$ Zeeman term acting in both the central region and leads, with conserved $\sigma_z$, the spin-dependent shift $s g\mu_BB/2$ is an exact rigid translation of the corresponding spin block. Direct NEGF evaluation at the shifted energies gives Table~\ref{tab:zeeman}. The resonance conductance can change noticeably, especially at point B, while the spin polarization remains essentially unchanged.

\begin{table*}[t]
\caption{Effect of a $g=2$ Zeeman scale. $G$ is in $\mu$S and $g\mu_BB$ in meV.}
\label{tab:zeeman}
\begin{ruledtabular}
\begin{tabular}{cccccc}
Point & $B$ (T) & $g\mu_BB$ & $G$ (no Zeeman) & $G$ ($g=2$) & $P_T$ ($g=2$)\\
\hline
A & 0.132 & 0.0152 & 38.84 & 38.70 & $-0.9922$\\
B & 0.263 & 0.0305 & 10.20 & 8.01 & $-0.9985$\\
E & 0.395 & 0.0457 & 36.44 & 34.93 & $-0.9956$\\
\end{tabular}
\end{ruledtabular}
\end{table*}

Thus the spin filter is not a conventional Zeeman filter: the altermagnetic/interferometric separation between the spin-resolved resonances is larger than the ordinary sub-tesla Zeeman scale, and the near-unity polarization survives when that uniform Zeeman shift is included exactly under the stated assumptions.

\section{Contact-matching sensitivity}
\label{app:contact}

The production geometry uses an abrupt opening of the Tan--Inkson confinement in the contact aperture, while the altermagnetic hopping extends through the finite neck up to the boundary with the ideal nonmagnetic lead self-energy. To assess whether the high polarization is purely an interface artifact, two alternative matching profiles were tested at the production mesh: (i) a smooth $10$~nm taper of the Tan--Inkson confinement, and (ii) the same potential taper combined with a smooth reduction of the altermagnetic anisotropy through the neck. Results are shown in Table~\ref{tab:contact}.

\begin{table*}[t]
\caption{Contact-matching test. $G$ is in $\mu$S.}
\label{tab:contact}
\begin{ruledtabular}
\begin{tabular}{ccccccc}
Point & \multicolumn{2}{c}{abrupt} & \multicolumn{2}{c}{smooth $V$} & \multicolumn{2}{c}{smooth $V$ and $\beta$}\\
 & $G$ & $P_T$ & $G$ & $P_T$ & $G$ & $P_T$\\
\hline
A & 38.84 & $-0.9919$ & 38.16 & $-0.9927$ & 27.08 & $-0.9989$\\
B & 10.20 & $-0.9990$ & 11.19 & $-0.9998$ & 2.53 & $-0.9399$\\
E & 36.44 & $-0.9944$ & 38.10 & $-0.9942$ & 19.53 & $-0.9829$\\
\end{tabular}
\end{ruledtabular}
\end{table*}

The absolute conductance is interface dependent, as expected for a coherent resonant device, but strong spin selectivity survives the smooth matching tests. Contact engineering is therefore part of the device physics rather than a numerically irrelevant detail.

\section{Dense Fourier extraction of the Aharonov--Bohm phase}
\label{app:dense_fourier}

The first AB harmonic was recomputed using 48 points per confined-flux period. This avoids aliasing from higher harmonics that are sizable in the coherent ring. At
\begin{equation}
B=(0,0.132,0.263,0.395)~\mathrm{T},
\end{equation}
the extracted unwrapped phases are
\begin{equation}
\delta_1/\pi=(-1.0000,-0.4828,0.0343,0.5513),
\end{equation}
and the corresponding amplitudes are
\begin{equation}
A_1=(9.33,9.33,9.34,9.36)~\mu\mathrm{S}.
\end{equation}
The amplitudes are nearly field independent. A linear fit to $\delta_1(B)$ gives
\begin{equation}
A_{\rm eff}=\frac{\Phi_0}{2\pi}\frac{d\delta_1}{dB}
=8.12\times10^{-15}\ {\rm m^2}
=1.0342\,\pi r_0^2,
\end{equation}
and therefore $\Delta B=\Phi_0/A_{\rm eff}=0.509$~T. After shifting the $B=0$ curve by $BA_{\rm eff}$, the mean absolute residual of the finite-$B$ conductance cuts is only $0.26$--$0.39~\mu$S over the sampled fields.

\section*{Data and code availability}
The data that support the findings of this study and the routines used to perform the calculations and generate the figures are available from the authors upon reasonable request.

\begin{acknowledgments}
The author acknowledges support from the Conselho Nacional de
Desenvolvimento Cient\'ifico e Tecnol\'ogico (CNPq), under Grant
No.~305427/2026-1; the Funda\c{c}\~ao de Amparo \`a Pesquisa e ao
Desenvolvimento Cient\'ifico e Tecnol\'ogico do Maranh\~ao (FAPEMA),
under Grant No.~UNIVERSAL-06395/22; and the Coordena\c{c}\~ao de
Aperfei\c{c}oamento de Pessoal de N\'ivel Superior (CAPES), Brazil,
under Finance Code 001.

\end{acknowledgments}

\section*{Declaration of AI-assisted tools}
AI-assisted tools were used as auxiliary resources during manuscript preparation and the analytical and numerical workflow. These included ChatGPT (OpenAI), with brief initial use of GPT Astra followed mainly by GPT Pro and GPT-5.6 Sol, and Claude Opus 5.5 (Anthropic). Their use encompassed language refinement, algebraic and numerical cross-checks, assistance with code development and debugging, and organization of the analysis. The author directed the workflow, reviewed the mathematical derivations and computational outputs, checked the physical interpretations and conclusions, and assumes full responsibility for the scientific content.


\begin{thebibliography}{99}

\bibitem{SmejkalPRX2022Phase}
L.~\v{S}mejkal, J.~Sinova, and T.~Jungwirth,
``Beyond Conventional Ferromagnetism and Antiferromagnetism: A Phase with Nonrelativistic Spin and Crystal Rotation Symmetry,''
Phys. Rev. X \textbf{12}, 031042 (2022),
\href{https://doi.org/10.1103/PhysRevX.12.031042}{doi:10.1103/PhysRevX.12.031042}.

\bibitem{SmejkalPRX2022Landscape}
L.~\v{S}mejkal, J.~Sinova, and T.~Jungwirth,
``Emerging Research Landscape of Altermagnetism,''
Phys. Rev. X \textbf{12}, 040501 (2022),
\href{https://doi.org/10.1103/PhysRevX.12.040501}{doi:10.1103/PhysRevX.12.040501}.

\bibitem{JungwirthNatPhys2026}
T.~Jungwirth, J.~Sinova, P.~Wadley, D.~Kriegner, H.~Reichlov\'a, F.~Krizek, H.~Ohno, \textit{et al.},
``Altermagnetic spintronics,''
Nat. Phys. \textbf{22}, 1012--1021 (2026),
\href{https://doi.org/10.1038/s41567-026-03337-w}{doi:10.1038/s41567-026-03337-w}.

\bibitem{SongNatRevMat2025}
C.~Song, H.~Bai, Z.~Zhou, \textit{et al.},
``Altermagnets as a new class of functional materials,''
Nat. Rev. Mater. \textbf{10}, 473--485 (2025),
\href{https://doi.org/10.1038/s41578-025-00779-1}{doi:10.1038/s41578-025-00779-1}.

\bibitem{JungwirthNature2026}
T.~Jungwirth, J.~Sinova, R.~M.~Fernandes, \textit{et al.},
``Symmetry, microscopy and spectroscopy signatures of altermagnetism,''
Nature \textbf{649}, 837--847 (2026),
\href{https://doi.org/10.1038/s41586-025-09883-2}{doi:10.1038/s41586-025-09883-2}.

\bibitem{McClartyRau2024}
P.~A.~McClarty and J.~G.~Rau,
``Landau Theory of Altermagnetism,''
Phys. Rev. Lett. \textbf{132}, 176702 (2024),
\href{https://doi.org/10.1103/PhysRevLett.132.176702}{doi:10.1103/PhysRevLett.132.176702}.

\bibitem{CheongClassification2025}
S.-W.~Cheong and F.-T.~Huang,
``Altermagnetism classification,''
npj Quantum Mater. \textbf{10}, 38 (2025),
\href{https://doi.org/10.1038/s41535-025-00756-5}{doi:10.1038/s41535-025-00756-5}.

\bibitem{KrempaskyNature2024}
J.~Krempask\'y, L.~\v{S}mejkal, S.~W.~D'Souza, \textit{et al.},
``Altermagnetic lifting of Kramers spin degeneracy,''
Nature \textbf{626}, 517--522 (2024),
\href{https://doi.org/10.1038/s41586-023-06907-7}{doi:10.1038/s41586-023-06907-7}.

\bibitem{HarikiPRL2024}
A.~Hariki, A.~Dal Din, O.~J.~Amin, \textit{et al.},
``X-Ray Magnetic Circular Dichroism in Altermagnetic $\alpha$-MnTe,''
Phys. Rev. Lett. \textbf{132}, 176701 (2024),
\href{https://doi.org/10.1103/PhysRevLett.132.176701}{doi:10.1103/PhysRevLett.132.176701}.

\bibitem{LeeMnTe2024}
S.~Lee, S.~Lee, S.~Jung, \textit{et al.},
``Broken Kramers Degeneracy in Altermagnetic MnTe,''
Phys. Rev. Lett. \textbf{132}, 036702 (2024),
\href{https://doi.org/10.1103/PhysRevLett.132.036702}{doi:10.1103/PhysRevLett.132.036702}.

\bibitem{AminMnTe2024}
O.~J.~Amin, A.~Dal Din, E.~Golias, \textit{et al.},
``Nanoscale imaging and control of altermagnetism in MnTe,''
Nature \textbf{636}, 348--353 (2024),
\href{https://doi.org/10.1038/s41586-024-08234-x}{doi:10.1038/s41586-024-08234-x}.

\bibitem{FengRuO22022}
Z.~Feng, X.~Zhou, L.~\v{S}mejkal, \textit{et al.},
``An anomalous Hall effect in altermagnetic ruthenium dioxide,''
Nat. Electron. \textbf{5}, 735--743 (2022),
\href{https://doi.org/10.1038/s41928-022-00866-z}{doi:10.1038/s41928-022-00866-z}.

\bibitem{ZhuPlaid2024}
Y.-P.~Zhu, X.~Chen, X.-R.~Liu, \textit{et al.},
``Observation of plaid-like spin splitting in a noncoplanar antiferromagnet,''
Nature \textbf{626}, 523--528 (2024),
\href{https://doi.org/10.1038/s41586-024-07023-w}{doi:10.1038/s41586-024-07023-w}.

\bibitem{GuoRuO22024}
Y.~Guo, J.~Zhang, Z.~Zhu, \textit{et al.},
``Direct and Inverse Spin Splitting Effects in Altermagnetic RuO$_2$,''
Adv. Sci. \textbf{11}, 2400967 (2024),
\href{https://doi.org/10.1002/advs.202400967}{doi:10.1002/advs.202400967}.

\bibitem{GonzalezMnTe2024}
R.~D.~Gonzalez Betancourt, J.~Zub\'a\v{c}, K.~Geishendorf, \textit{et al.},
``Anisotropic magnetoresistance in altermagnetic MnTe,''
npj Spintronics \textbf{2}, 45 (2024),
\href{https://doi.org/10.1038/s44306-024-00046-z}{doi:10.1038/s44306-024-00046-z}.

\bibitem{SunPRB2023}
C.~Sun, A.~Brataas, and J.~Linder,
``Andreev reflection in altermagnets,''
Phys. Rev. B \textbf{108}, 054511 (2023),
\href{https://doi.org/10.1103/PhysRevB.108.054511}{doi:10.1103/PhysRevB.108.054511}.

\bibitem{ZarzuelaPRB2025}
R.~Zarzuela, R.~Jaeschke-Ubiergo, O.~Gomonay, L.~\v{S}mejkal, and J.~Sinova,
``Transport theory and spin-transfer physics in $d$-wave altermagnets,''
Phys. Rev. B \textbf{111}, 064422 (2025),
\href{https://doi.org/10.1103/PhysRevB.111.064422}{doi:10.1103/PhysRevB.111.064422};
Erratum: Phys. Rev. B \textbf{111}, 219901 (2025),
\href{https://doi.org/10.1103/gw39-v17c}{doi:10.1103/gw39-v17c}.

\bibitem{ZhangPRL2025}
X.-P.~Zhang, W.~Feng, R.-W.~Zhang, X.~Fan, X.~Wang, and Y.~Yao,
``Theory of Anisotropic Magnetoresistance in Altermagnets and Its Applications,''
Phys. Rev. Lett. \textbf{135}, 266706 (2025),
\href{https://doi.org/10.1103/32zc-ggjy}{doi:10.1103/32zc-ggjy}.

\bibitem{SooriPRM2026}
A.~Soori,
``Geometry induced net spin polarization of $d$-wave altermagnets,''
Phys. Rev. Materials \textbf{10}, 044406 (2026),
\href{https://doi.org/10.1103/8ntv-rwyn}{doi:10.1103/8ntv-rwyn}.

\bibitem{GomonayTextures2024}
O.~Gomonay, V.~P.~Kravchuk, R.~Jaeschke-Ubiergo, \textit{et al.},
``Structure, control, and dynamics of altermagnetic textures,''
npj Spintronics \textbf{2}, 35 (2024),
\href{https://doi.org/10.1038/s44306-024-00042-3}{doi:10.1038/s44306-024-00042-3}.

\bibitem{SatoAHE2024}
T.~Sato, S.~Haddad, I.~C.~Fulga, F.~F.~Assaad, and J.~van den Brink,
``Altermagnetic Anomalous Hall Effect Emerging from Electronic Correlations,''
Phys. Rev. Lett. \textbf{133}, 086503 (2024),
\href{https://doi.org/10.1103/PhysRevLett.133.086503}{doi:10.1103/PhysRevLett.133.086503}.

\bibitem{FangQuantumGeometry2024}
Y.~Fang, J.~Cano, and S.~A.~A.~Ghorashi,
``Quantum Geometry Induced Nonlinear Transport in Altermagnets,''
Phys. Rev. Lett. \textbf{133}, 106701 (2024),
\href{https://doi.org/10.1103/PhysRevLett.133.106701}{doi:10.1103/PhysRevLett.133.106701}.

\bibitem{JaeschkeSupercell2024}
R.~Jaeschke-Ubiergo, V.~K.~Bharadwaj, T.~Jungwirth, L.~\v{S}mejkal, and J.~Sinova,
``Supercell altermagnets,''
Phys. Rev. B \textbf{109}, 094425 (2024),
\href{https://doi.org/10.1103/PhysRevB.109.094425}{doi:10.1103/PhysRevB.109.094425}.

\bibitem{ZhangPairing2024}
S.-B.~Zhang, L.-H.~Hu, and T.~Neupert,
``Finite-momentum Cooper pairing in proximitized altermagnets,''
Nat. Commun. \textbf{15}, 1801 (2024),
\href{https://doi.org/10.1038/s41467-024-45951-3}{doi:10.1038/s41467-024-45951-3}.

\bibitem{AharonovBohm1959}
Y.~Aharonov and D.~Bohm,
``Significance of Electromagnetic Potentials in the Quantum Theory,''
Phys. Rev. \textbf{115}, 485--491 (1959),
\href{https://doi.org/10.1103/PhysRev.115.485}{doi:10.1103/PhysRev.115.485}.

\bibitem{TanInkson1996}
W.-C.~Tan and J.~C.~Inkson,
``Landau quantization and the Aharonov--Bohm effect in a two-dimensional ring,''
Phys. Rev. B \textbf{53}, 6947--6950 (1996),
\href{https://doi.org/10.1103/PhysRevB.53.6947}{doi:10.1103/PhysRevB.53.6947}.

\bibitem{ButtikerImryLandauer1985}
M.~B\"uttiker, Y.~Imry, R.~Landauer, and S.~Pinhas,
``Generalized many-channel conductance formula with application to small rings,''
Phys. Rev. B \textbf{31}, 6207--6215 (1985),
\href{https://doi.org/10.1103/PhysRevB.31.6207}{doi:10.1103/PhysRevB.31.6207}.

\bibitem{TanInkson1999}
W.-C.~Tan and J.~C.~Inkson,
``Magnetization, persistent currents, and their relation in quantum rings and dots,''
Phys. Rev. B \textbf{60}, 5626--5635 (1999),
\href{https://doi.org/10.1103/PhysRevB.60.5626}{doi:10.1103/PhysRevB.60.5626}.

\bibitem{YeyatiButtiker1995}
A.~Levy Yeyati and M.~B\"uttiker,
``Aharonov--Bohm oscillations in a mesoscopic ring with a quantum dot,''
Phys. Rev. B \textbf{52}, R14360--R14363 (1995),
\href{https://doi.org/10.1103/PhysRevB.52.R14360}{doi:10.1103/PhysRevB.52.R14360}.

\bibitem{HackenbroichWeidenmuller1996}
G.~Hackenbroich and H.~A.~Weidenm\"uller,
``Conductance through a quantum dot in an Aharonov--Bohm ring,''
Phys. Rev. B \textbf{53}, 16379--16390 (1996),
\href{https://doi.org/10.1103/PhysRevB.53.16379}{doi:10.1103/PhysRevB.53.16379}.

\bibitem{Cernicchiaro1997}
G.~Cernicchiaro, T.~Martin, K.~Hasselbach, D.~Mailly, and A.~Benoit,
``Channel Interference in a Quasiballistic Aharonov--Bohm Experiment,''
Phys. Rev. Lett. \textbf{79}, 273--276 (1997),
\href{https://doi.org/10.1103/PhysRevLett.79.273}{doi:10.1103/PhysRevLett.79.273}.

\bibitem{Aharony2006}
A.~Aharony, O.~Entin-Wohlman, T.~Otsuka, \textit{et al.},
``Breakdown of phase rigidity and variations of the Fano effect in closed Aharonov--Bohm interferometers,''
Phys. Rev. B \textbf{73}, 195329 (2006),
\href{https://doi.org/10.1103/PhysRevB.73.195329}{doi:10.1103/PhysRevB.73.195329}.

\bibitem{Fuhrer2006}
A.~Fuhrer, P.~Brusheim, T.~Ihn, \textit{et al.},
``Fano effect in a quantum-ring--quantum-dot system with tunable coupling,''
Phys. Rev. B \textbf{73}, 205326 (2006),
\href{https://doi.org/10.1103/PhysRevB.73.205326}{doi:10.1103/PhysRevB.73.205326}.

\bibitem{Nakanishi2004}
T.~Nakanishi, K.~Terakura, and T.~Ando,
``Theory of Fano effects in an Aharonov--Bohm ring with a quantum dot,''
Phys. Rev. B \textbf{69}, 115307 (2004),
\href{https://doi.org/10.1103/PhysRevB.69.115307}{doi:10.1103/PhysRevB.69.115307}.

\bibitem{Moldoveanu2005}
V.~Moldoveanu, M.~\c{T}olea, A.~Aldea, and B.~Tanatar,
``Resonant and coherent transport through Aharonov--Bohm interferometers with coupled quantum dots,''
Phys. Rev. B \textbf{71}, 125338 (2005),
\href{https://doi.org/10.1103/PhysRevB.71.125338}{doi:10.1103/PhysRevB.71.125338}.

\bibitem{Vargiamidis2006}
V.~Vargiamidis and H.~M.~Polatoglou,
``Fano resonance and persistent current in mesoscopic open rings: Influence of coupling and Aharonov--Bohm flux,''
Phys. Rev. B \textbf{74}, 235323 (2006),
\href{https://doi.org/10.1103/PhysRevB.74.235323}{doi:10.1103/PhysRevB.74.235323}.

\bibitem{FrustagliaRichter2004}
D.~Frustaglia and K.~Richter,
``Spin interference effects in ring conductors subject to Rashba coupling,''
Phys. Rev. B \textbf{69}, 235310 (2004),
\href{https://doi.org/10.1103/PhysRevB.69.235310}{doi:10.1103/PhysRevB.69.235310}.

\bibitem{Cheung1989}
H.-F.~Cheung, E.~K.~Riedel, and Y.~Gefen,
``Persistent Currents in Mesoscopic Rings and Cylinders,''
Phys. Rev. Lett. \textbf{62}, 587--590 (1989),
\href{https://doi.org/10.1103/PhysRevLett.62.587}{doi:10.1103/PhysRevLett.62.587}.

\bibitem{Levy1990}
L.~P.~L\'evy, G.~Dolan, J.~Dunsmuir, and H.~Bouchiat,
``Magnetization of mesoscopic copper rings: Evidence for persistent currents,''
Phys. Rev. Lett. \textbf{64}, 2074--2077 (1990),
\href{https://doi.org/10.1103/PhysRevLett.64.2074}{doi:10.1103/PhysRevLett.64.2074}.

\bibitem{AmbegaokarEckern1990}
V.~Ambegaokar and U.~Eckern,
``Coherence and persistent currents in mesoscopic rings,''
Phys. Rev. Lett. \textbf{65}, 381--384 (1990),
\href{https://doi.org/10.1103/PhysRevLett.65.381}{doi:10.1103/PhysRevLett.65.381}.

\bibitem{Montambaux1990}
G.~Montambaux, H.~Bouchiat, D.~Sigeti, and R.~Friesner,
``Persistent currents in mesoscopic metallic rings: Ensemble average,''
Phys. Rev. B \textbf{42}, 7647--7650 (1990),
\href{https://doi.org/10.1103/PhysRevB.42.7647}{doi:10.1103/PhysRevB.42.7647}.

\bibitem{vonOppen1991}
F.~von Oppen and E.~K.~Riedel,
``Average persistent current in a mesoscopic ring,''
Phys. Rev. Lett. \textbf{66}, 84--87 (1991),
\href{https://doi.org/10.1103/PhysRevLett.66.84}{doi:10.1103/PhysRevLett.66.84}.

\bibitem{Loss1990}
D.~Loss, P.~Goldbart, and A.~V.~Balatsky,
``Berry's phase and persistent charge and spin currents in textured mesoscopic rings,''
Phys. Rev. Lett. \textbf{65}, 1655--1658 (1990),
\href{https://doi.org/10.1103/PhysRevLett.65.1655}{doi:10.1103/PhysRevLett.65.1655}.

\bibitem{ZhangNatPhys2025RbV}
F.~Zhang, X.~Cheng, Z.~Yin, \textit{et al.},
``Crystal-symmetry-paired spin--valley locking in a layered room-temperature metallic altermagnet candidate,''
Nat. Phys. \textbf{21}, 760--767 (2025),
\href{https://doi.org/10.1038/s41567-025-02864-2}{doi:10.1038/s41567-025-02864-2}.

\bibitem{HuPRX2025RbV}
M.~Hu, X.~Cheng, Z.~Huang, and J.~Liu,
``Catalog of C-Paired Spin-Momentum Locking in Antiferromagnetic Systems,''
Phys. Rev. X \textbf{15}, 021083 (2025),
\href{https://doi.org/10.1103/PhysRevX.15.021083}{doi:10.1103/PhysRevX.15.021083}.

\bibitem{MaoPRB2026RbV}
C.~Mao, S.~Liu, S.~Li, J.~Yang, and J.~Yang,
``Interfacial bond tailored altermagnetic tunnel junctions,''
Phys. Rev. B \textbf{113}, 174409 (2026),
\href{https://doi.org/10.1103/xcwv-bvkr}{doi:10.1103/xcwv-bvkr}.

\bibitem{XieCPB2026RbV}
W.~Xie, C.~Liu, F.~Zhang, \textit{et al.},
``G-type antiferromagnetic structure in the metallic $d$-wave altermagnet $\mathrm{Rb}_{1-\delta}\mathrm{V}_2\mathrm{Te}_2\mathrm{O}$,''
Chin. Phys. B \textbf{35}, 087503 (2026),
\href{https://doi.org/10.1088/1674-1056/ae8f1a}{doi:10.1088/1674-1056/ae8f1a}.

\bibitem{Caroli1971}
C.~Caroli, R.~Combescot, P.~Nozi\`eres, and D.~Saint-James,
``Direct calculation of the tunneling current,''
J. Phys. C: Solid State Phys. \textbf{4}, 916--929 (1971),
\href{https://doi.org/10.1088/0022-3719/4/8/018}{doi:10.1088/0022-3719/4/8/018}.

\bibitem{Datta1995}
S.~Datta, \textit{Electronic Transport in Mesoscopic Systems}
(Cambridge University Press, Cambridge, 1995),
\href{https://doi.org/10.1017/CBO9780511805776}{doi:10.1017/CBO9780511805776}.

\bibitem{Onsager1931}
L.~Onsager,
``Reciprocal Relations in Irreversible Processes. I,''
Phys. Rev. \textbf{37}, 405--426 (1931),
\href{https://doi.org/10.1103/PhysRev.37.405}{doi:10.1103/PhysRev.37.405};
``Reciprocal Relations in Irreversible Processes. II,''
Phys. Rev. \textbf{38}, 2265--2279 (1931),
\href{https://doi.org/10.1103/PhysRev.38.2265}{doi:10.1103/PhysRev.38.2265}.

\bibitem{Casimir1945}
H.~B.~G.~Casimir,
``On Onsager's Principle of Microscopic Reversibility,''
Rev. Mod. Phys. \textbf{17}, 343--350 (1945),
\href{https://doi.org/10.1103/RevModPhys.17.343}{doi:10.1103/RevModPhys.17.343}.

\bibitem{Buttiker1986}
M.~B\"uttiker,
``Four-Terminal Phase-Coherent Conductance,''
Phys. Rev. Lett. \textbf{57}, 1761--1764 (1986),
\href{https://doi.org/10.1103/PhysRevLett.57.1761}{doi:10.1103/PhysRevLett.57.1761}.

\bibitem{YacobyPRB1996}
A.~Yacoby, R.~Schuster, and M.~Heiblum,
``Phase rigidity and $h/2e$ oscillations in a single-ring Aharonov--Bohm experiment,''
Phys. Rev. B \textbf{53}, 9583--9586 (1996),
\href{https://doi.org/10.1103/PhysRevB.53.9583}{doi:10.1103/PhysRevB.53.9583}.

\bibitem{BarangerStone1989}
H.~U.~Baranger and A.~D.~Stone,
``Electrical linear-response theory in an arbitrary magnetic field: A new Fermi-surface formation,''
Phys. Rev. B \textbf{40}, 8169--8193 (1989),
\href{https://doi.org/10.1103/PhysRevB.40.8169}{doi:10.1103/PhysRevB.40.8169}.

\bibitem{Hofstadter1976}
D.~R.~Hofstadter,
``Energy levels and wave functions of Bloch electrons in rational and irrational magnetic fields,''
Phys. Rev. B \textbf{14}, 2239--2249 (1976),
\href{https://doi.org/10.1103/PhysRevB.14.2239}{doi:10.1103/PhysRevB.14.2239}.

\bibitem{DoPeierls2022}
T.-N.~Do, P.-H.~Shih, H.~Lin, D.~Huang, G.~Gumbs, and T.-R.~Chang,
``Generalized Peierls substitution for the tight-binding model of twisted graphene systems in a magnetic field,''
Phys. Rev. B \textbf{105}, 235418 (2022),
\href{https://doi.org/10.1103/PhysRevB.105.235418}{doi:10.1103/PhysRevB.105.235418}.

\bibitem{FisherLee1981}
D.~S.~Fisher and P.~A.~Lee,
``Relation between conductivity and transmission matrix,''
Phys. Rev. B \textbf{23}, 6851--6854 (1981),
\href{https://doi.org/10.1103/PhysRevB.23.6851}{doi:10.1103/PhysRevB.23.6851}.

\bibitem{Beenakker1997}
C.~W.~J.~Beenakker,
``Random-matrix theory of quantum transport,''
Rev. Mod. Phys. \textbf{69}, 731--808 (1997),
\href{https://doi.org/10.1103/RevModPhys.69.731}{doi:10.1103/RevModPhys.69.731}.

\bibitem{ButtikerPRB1986Coherence}
M.~B\"uttiker,
``Role of quantum coherence in series resistors,''
Phys. Rev. B \textbf{33}, 3020--3026 (1986),
\href{https://doi.org/10.1103/PhysRevB.33.3020}{doi:10.1103/PhysRevB.33.3020}.

\bibitem{DAmatoPastawski1990}
J.~L.~D'Amato and H.~M.~Pastawski,
``Conductance of a disordered linear chain including inelastic scattering events,''
Phys. Rev. B \textbf{41}, 7411--7420 (1990),
\href{https://doi.org/10.1103/PhysRevB.41.7411}{doi:10.1103/PhysRevB.41.7411}.

\bibitem{Forster2007}
H.~F\"orster, P.~Samuelsson, S.~Pilgram, and M.~B\"uttiker,
``Voltage and dephasing probes in mesoscopic conductors: A study of full-counting statistics,''
Phys. Rev. B \textbf{75}, 035340 (2007),
\href{https://doi.org/10.1103/PhysRevB.75.035340}{doi:10.1103/PhysRevB.75.035340}.

\bibitem{SivanImry1986}
U.~Sivan and Y.~Imry,
``Multichannel Landauer formula for thermoelectric transport with application to thermopower near the mobility edge,''
Phys. Rev. B \textbf{33}, 551--558 (1986),
\href{https://doi.org/10.1103/PhysRevB.33.551}{doi:10.1103/PhysRevB.33.551}.

\bibitem{LopezSancho1985}
M.~P.~L\'opez Sancho, J.~M.~L\'opez Sancho, and J.~Rubio,
``Highly convergent schemes for the calculation of bulk and surface Green functions,''
J. Phys. F: Met. Phys. \textbf{15}, 851--858 (1985),
\href{https://doi.org/10.1088/0305-4608/15/4/009}{doi:10.1088/0305-4608/15/4/009}.

\bibitem{MeirWingreen1992}
Y.~Meir and N.~S.~Wingreen,
``Landauer formula for the current through an interacting electron region,''
Phys. Rev. Lett. \textbf{68}, 2512--2515 (1992),
\href{https://doi.org/10.1103/PhysRevLett.68.2512}{doi:10.1103/PhysRevLett.68.2512}.

\bibitem{Peierls1933}
R.~Peierls,
``Zur Theorie des Diamagnetismus von Leitungselektronen,''
Z. Phys. \textbf{80}, 763--791 (1933),
\href{https://doi.org/10.1007/BF01342591}{doi:10.1007/BF01342591}.

\bibitem{Anderson1958}
P.~W.~Anderson,
``Absence of Diffusion in Certain Random Lattices,''
Phys. Rev. \textbf{109}, 1492--1505 (1958),
\href{https://doi.org/10.1103/PhysRev.109.1492}{doi:10.1103/PhysRev.109.1492}.

\bibitem{LeeStone1985}
P.~A.~Lee and A.~D.~Stone,
``Universal Conductance Fluctuations in Metals,''
Phys. Rev. Lett. \textbf{55}, 1622--1625 (1985),
\href{https://doi.org/10.1103/PhysRevLett.55.1622}{doi:10.1103/PhysRevLett.55.1622}.

\bibitem{LeeStoneFukuyama1987}
P.~A.~Lee, A.~D.~Stone, and H.~Fukuyama,
``Universal conductance fluctuations in metals: Effects of finite temperature, interactions, and magnetic field,''
Phys. Rev. B \textbf{35}, 1039--1070 (1987),
\href{https://doi.org/10.1103/PhysRevB.35.1039}{doi:10.1103/PhysRevB.35.1039}.

\end{thebibliography}
\end{document}